\documentclass[a4paper]{article}

\usepackage{amsmath, amssymb}
\usepackage{booktabs}
\usepackage{graphicx}
\usepackage[dvipsnames]{xcolor}
\usepackage{filemod}
\usepackage{siunitx}
\usepackage{mathtools}
\usepackage{multirow}
\usepackage{algorithm}
\usepackage{adjustbox}
\usepackage{tabularx}
\usepackage[noend]{algpseudocode}
\usepackage{url}
\usepackage[hidelinks]{hyperref}
\usepackage[margin = 2cm]{geometry}

\usepackage{tikz}
\usepackage{pgfplots}
\usepgfplotslibrary{groupplots}
\pgfplotsset{compat=1.17,
    legend image code/.code={
    \draw[]
    plot coordinates {
    (0cm,0cm)
    (0.3cm,0cm)
    };%
}}
\pgfplotsset{every x tick label/.append style={font=\footnotesize}}
\pgfplotsset{every y tick label/.append style={font=\footnotesize}}
\pgfplotsset{every y label/.append style={font=\footnotesize}}
\pgfplotsset{every x label/.append style={font=\footnotesize}}

\usepgfplotslibrary{external} 
\newlength{\figurewidth}
\newlength{\figureheight}

\newcommand{\real}{\ensuremath{\mathbb R}}
\newcommand{\sat}{\ensuremath{\mathrm{sat}}}
\newcommand{\norm}[1]{\left\lVert#1\right\rVert}

\newcommand{\dz}{\ensuremath{\mathrm{dz}}}
\renewcommand\exp[1]{\ensuremath{\mathrm{e}^{#1}}}

\newcommand\smallmat[1]{\left[\begin{smallmatrix}#1\end{smallmatrix}\right]}
\newcommand\bigmat[1]{\begin{bmatrix}#1\end{bmatrix}}

\newcommand\blfootnote[1]{%
  \begingroup
  \renewcommand\thefootnote{}\footnote{#1}%
  \addtocounter{footnote}{-1}%
  \endgroup
}

\graphicspath{{./fig/}}

\newtheorem{lemma}{Lemma}

\allowdisplaybreaks

\newtheorem{proposition}{Proposition}

\newtheorem{thm}{Theorem}

\title{Hybrid Lyapunov-based feedback stabilization of bipedal locomotion based on reference spreading}
\author{Riccardo Bertollo$^1$, Gianni Lunardi$^2$, Andrea Del Prete$^2$, Luca Zaccarian$^{2,3}$}
\date{\small $^1$Dept. of Mechanical Engineering, Eindhoven University of Technology (The Netherlands) \\
$^2$Dept. of Industrial Engineering, University of Trento (Italy) \\
$^3$LAAS-CNRS, Universit{\'e} de Toulouse, CNRS Toulouse (France)}

\newenvironment{proof}{\smallskip \noindent{\textbf{Proof:}}}{\hfill \hspace*{1pt}\hfill $\square$}

\begin{document}
    
    \maketitle



    \begin{abstract}
        We propose a hybrid formulation of the linear inverted pendulum model for bipedal locomotion, where the foot switches are triggered based on the center of mass position, removing the need for pre-defined footstep timings.
        Using a concept similar to reference spreading, we define nontrivial tracking error coordinates induced by our hybrid model. These coordinates enjoy desirable linear flow dynamics and rather elegant jump dynamics perturbed by a suitable extended class ${\mathcal K}_\infty$ function of the position error.
        We stabilize this hybrid error dynamics using a saturated feedback controller, selecting its gains by solving a convex optimization problem.
        We prove local asymptotic stability of the tracking error and provide a certified estimate of the basin of attraction, comparing it with a numerical estimate obtained from the integration of the closed-loop dynamics.
        Simulations on a full-body model of a real robot show the practical applicability of the proposed framework and its advantages with respect to a standard model predictive control formulation.
    \end{abstract}

\section{Introduction}
    \blfootnote{The research is supported in part by the MIUR in the 2020 PRIN framework under Grant number 2020RTWES4 (DOCEAT).}
    Legged robots offer ample mobility potential over rough terrain by decoupling the environment and the main body of the robot through the use of articulated legs.
    This allows the motion of the main body to be independent of the roughness of the terrain and allows the legs to temporarily leave contact with the ground to overcome isolated footholds.
    Compared to standard wheeled robots, this leads to improved performance potential, but comes at the cost of highly increased complexity.
    Indeed, while the first appearance of digitally controlled legged robots dates back to the late 1960s (for a historical overview, see \cite{Wieber2016}), only in the early 2000s the real-life capabilities of these robots began to resemble the animal behavior.
    
    One of the main sources of complexity in legged robots is the related multi-body dynamics, which is high-dimensional and nonlinear.
    Moreover, the intermittent contacts with the environment are typically modeled as infinitely rigid, which makes the dynamics hybrid.
    Finally, these systems are highly constrained: joint positions, velocities and torques are bounded, and contact forces are limited by friction cones \cite{RoboticSurvey}.
    The most common approaches to deal with these systems rely on a pre-specification of the contact phases (i.e. which contacts are active at any time instant), which transforms the dynamics from hybrid to time-varying. 
    After this transformation, standard trajectory optimization techniques can be used for generating locomotion behaviors accounting for the system constraints.
    However, the resulting optimization problems are still nonconvex and high-dimensional.

    To further reduce the model complexity and simplify the control design, researchers have proposed over the years different reduced-order models, which capture the dominant robot dynamics with a reduced state size \cite{RoboticSurvey}.
    The most well-known example of this is the Linear Inverted Pendulum Model (LIPM) \cite{KajitaICRA03}, which considers position and velocity of the center of mass (CoM) of the robot and the contact locations, neglecting joint-space constraints, angular momentum, inertias and variable CoM height.

    Given the simple linear nature of the LIPM, optimal control strategies can be easily applied to compute the optimal CoM motion and foot placement \cite{HerdtAR10}.
    Assuming that the times of the steps are known, the OCP solution provides the positions of the feet on the ground, and this allows modeling the robot as a continuous-time time-invariant linear system, with time-varying bounds on the control input (i.e. the center of pressure, a.k.a. Zero Moment Point).
    However, since the contact timings are supposed to be known in advance, this framework is unsuited to analyze what would happen if the contact timings differed from the expected ones, which, to a certain extent, is always the case in practice.
    Another option is to include the duration of the contact phase as decision variables in the optimization, as in \cite{PontonICRA18}; however, this has been shown to make the OCP nonconvex and hard to solve in a model-predictive-control scheme.

When using the full multibody dynamics of the robot, and describing the contact transitions with a hybrid approach, numerous valuable local stabilization results have been produced by leveraging control Lyapunov functions (see, e.g., \cite{GallowayAccess15,AmesACC18,AmesRAL21} and references therein) or Lyapunov-based barrier functions \cite{RinconTSMCS22}, 
often in combination with the online solution of optimization problems (see, e.g., the quadratic programs in \cite{RinconTSMCS22,AmesACC20}). 
The strategy in \cite{AmesACC18} follows a well-established nonlinear control paradigm wherein an authentically hybrid dynamical behavior is modeled via automata that perform state-dependent switching between different operating modes (one or two contact feet). The corresponding control solutions (see, e.g., \cite{AmesTAC14,AmesNAHS17,SidorovNAHS19,MorrisTAC09} and references therein) are based on local solutions of the challenging stabilization problem of a hybrid limit cycle, often addressed by using suitable Poincar\'e maps and analyzing the ensuing hybrid zero dynamics, often associated with suitable input-to-state stability properties for increased robustness. Despite the large amount of literature on the subject, many important challenges are still unresolved behind this Lyapunov-based approach to the locomotion problem (see the extensive discussion in the broad survey \cite{GrizzleAuto14}).

    In this paper, we propose a novel formulation for the tracking of a periodic trajectory with a bipedal walking robot, based on the use of a hybrid dynamical systems framework \cite{TeelBook12}.
    Contrary to the classical approach, where the contact timings are fixed a priori, we assume that contacts occur when the CoM reaches a given position, using hybrid jumps to shift the reference frame at each step, so that it corresponds to the center of the foot on the ground. 
    To suitably deal with the non-synchronized contact times of the reference motion and the actual robot, we adopt a hybrid modification of the reference dynamics similar in nature to the so-called ``reference spreading'' solution discussed in \cite{VanSteenACC23,BiemondTAC19,ForniTAC13} and references therein, to handle impacting systems. Interestingly, the reference spreading paradigm has been applied to legged robots in \cite{SacconICRA17} with very encouraging results. Our generalization of this idea stems from the fact that only the velocity jumps in those impacting systems, whereas the position remains constant. Instead, the essential stabilization problem stemming from our LIPM-based approach witnesses a jump of the position variable (due to the shift of the reference frame at the center of the contact foot) and continuity of the velocity variable. This reversed scenario requires novel developments as compared to the existing literature and leads us to the definition of a non-trivial hybrid error coordinate depending on suitable timers augmenting the standard LIPM equations.
    
    Our solution presents two main advantages with respect to the classical fixed-contact approach.
    First, it well describes the natural interplay between the stepping frequency and the walking speed. As the robot walks faster, its stepping frequency automatically increases. As empirically shown in our results in Section~\ref{sec:full_body_sim}, this can allow the robot to keep walking in cases where a classic approach would lead to a fall.
    Secondly, in view of a well-posed formulation, it allows proving rigorous robust stability guarantees (see \cite[Chapter 7]{TeelBook12}).
    In particular, our stability proof is based on an optimized selection of a quadratic Lyapunov function for which we can ensure asymptotic stability of the error dynamics despite a mismatch in the impact times of the reference and actual motion. In addition, we provide optimality-based convex design conditions for the stabilizing gains, through a linear matrix inequality whose feasibility is proven to hold under an intuitive necessary and sufficient condition.

    In this first work, we focus on tracking a reference for the longitudinal motion of the robot's center of mass. However, through a suitable extended simulation environment we illustrate our results by way of full-body simulations on the humanoid robot Romeo.

    The remainder of the paper is organized as follows.
    In Section~\ref{sec:model} we present our novel hybrid formulation of the LIPM dynamics and the reference trajectory, also deriving a hybrid dynamics for the tracking error.
    In Section~\ref{sec:controller}, we propose a convex numerical optimization problem to select the gains of a saturated linear feedback controller and the parameters of a nonlinear hybrid Lyapunov function certifying local asymptotic stability of the error dynamics, with a certified estimate of the domain of attraction.
    In Section~\ref{sec:proof}, we prove our main result, namely local asymptotic stability of the trajectory tracking error.
    In Section~\ref{sec:simulations}, we illustrate our theoretical results through numerical simulations.
    In Section~\ref{sec:full_body_sim}, we present an algorithm for the implementation of the proposed control strategy on a full-body model of a biped, and report the simulation results on a 37-degree-of-freedom humanoid robot.

    \begin{figure*}
        \centering
        \includegraphics[width=0.95 \textwidth]{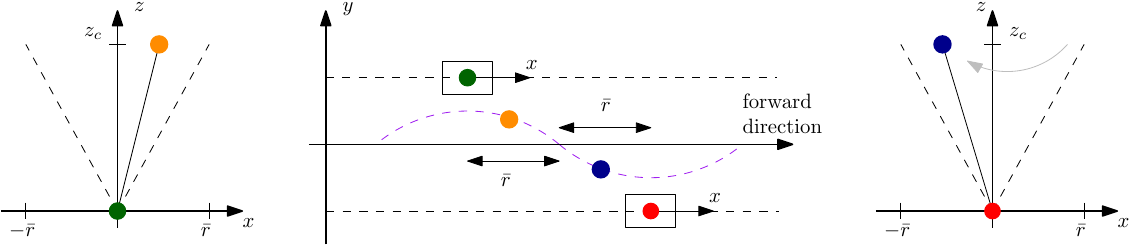}
        \caption{Scheme of the hybrid linear inverted pendulum model.}
        \label{fig:lipm}
    \end{figure*}

\section{Hybrid model}
\label{sec:model}
    \subsection{Biped longitudinal dynamics hybrid formulation}
        The LIPM is a reduced dynamical model of a walking robot that neglects the joint behaviors, but captures well the relationship between the contact forces and the CoM motion.
        In particular, the full-robot dynamics can be reduced to the LIPM under the following assumptions \cite{PontonICRA18}: {\it i)} the CoM height is constant, {\it ii)} the ground is flat, {\it iii)} the angular momentum about the CoM is null, {\it iv)} the contact friction is sufficiently high to avoid slippage.
        In the LIPM, the net effect of the contact forces is lumped in the center of pressure, which can be defined as a weighted average of the 
        contact points, using their associated vertical force as weight.
        For our analysis, we consider the projection of the center of mass on the longitudinal axis $x_c$, which obeys the following continuous-time dynamics
        \begin{align}
            \label{eq:com_ode}
            \Ddot x_c = \omega^2 ( x_c - u ) := \frac{g}{z_c} ( x_c - u ),
        \end{align}
        where $g$ is the gravity acceleration and the input $u \in \real$ represents the position of the center of pressure.
        Note that $u$ has by definition an upper and a lower bound, given by the foot length.
        
        We can represent dynamics \eqref{eq:com_ode}, by introducing the state variable 
        \begin{align}
            \label{eq:state}
            x := \left( x_p, x_v \right) := \left( x_c, \; \dot x_c \right),
        \end{align}
        whose continuous dynamics corresponds to
        \begin{align}
            \label{eq:com_ss}
            \begin{split}
                \dot x
                &= A x
                + B u 
                :=
                \begin{bmatrix}
                    0 & 1 \\ \omega^2 & 0
                \end{bmatrix} x
                +
                \begin{bmatrix}
                    0 \\ - \omega^2
                \end{bmatrix} u.
            \end{split}
        \end{align}
        In the classical continuous-time version of the LIPM, the reference frame is fixed, resulting in an unbounded value of the forward position $x_c$.
        Our model follows the continuous evolution described in \eqref{eq:com_ss}, but at every step we shift the reference frame to the center of the new support foot (under the typical assumption of no double support phase). For instance, this is shown in Figure~\ref{fig:lipm}, 
        where the origin of the frame is shifted from the green dot to the red dot when switching from the left foot contact to the right foot contact. In this way, the origin always corresponds to the center of the support foot.

        To suitably represent this mechanism via a hybrid model, we introduce the constant half-step size $\bar r$, corresponding to half of the longitudinal length of a step, in the absence of the control input and external perturbations.
        Then, the position variable $x_p = x_c$ is constrained to evolve in the compact set $[-\bar r, \bar r]$.
        With this overall state representation, the evolution of 
        the system state $x$ happens in the following set
        \begin{align}
            \mathcal X &:= [-\bar r, \bar r] \times \mathbb{R}.
        \end{align}
        According to the hybrid dynamical systems framework described in \cite{TeelBook12}, the state space is divided into a flow set $\mathcal C$ and a jump set $\mathcal D$, selected as follows
        \begin{align}
        \label{eq:flow_jump_set}
            \mathcal C:= \mathcal X, \quad\quad \mathcal D :=\{x \in \mathcal X \; : \; x_p = \bar r \},
        \end{align}
        %
        The corresponding hybrid dynamics is the following
        \begin{subequations}
        \label{eq:robot_plant}
            \begin{align}
            &\dot x = A x + B u & x \in \mathcal C,
            \label{eq:flow_law} \\
            &x^+ = x + F := x +  \bigmat{
                -2 \overline r \\ 0}, & x \in \mathcal D ,
            \label{eq:jump_law}
            \end{align}
        \end{subequations}
        Note that, strictly speaking, the center of mass velocity $x_v = \dot x_c$ could be negative.
        Thus, since the proposed hybrid dynamics does not contemplate jumps from $-\bar r$ to $\bar r$ (i.e., backwards steps), not all the solutions to this system are complete, meaning that, according to the definition in \cite[Chapter 2]{TeelBook12}, they may exit $\mathcal X = \mathcal C \cup \mathcal D$ and stop evolving.
        However, we still consider the result that we present to be of interest.
        Indeed, the reference trajectory has positive velocity, so for a small enough initial tracking error, in view of a saturated control action, the robot velocity never becomes negative.
        A deeper investigation on the matter, with the goal to obtain explicit characterization of the subset from which all solutions are complete, could be an interesting objective of future work.

\subsection{Reference signal}
\label{sec:reference}

The reference trajectory $(t,j) \mapsto x_r(t,j)$ is a symmetric periodic motion with period $T$ arising when fixing the center of pressure 
at the geometrical center of the foot touching the ground (namely, setting $u=0$ in \eqref{eq:com_ss}). 
In particular, the period $T$ depends on two free parameters: the half-step size $\bar r$, introduced above, and the peak forward speed $\bar v$, reached at the step change. 

\begin{figure}[!hbt]
    \centering
    \includegraphics{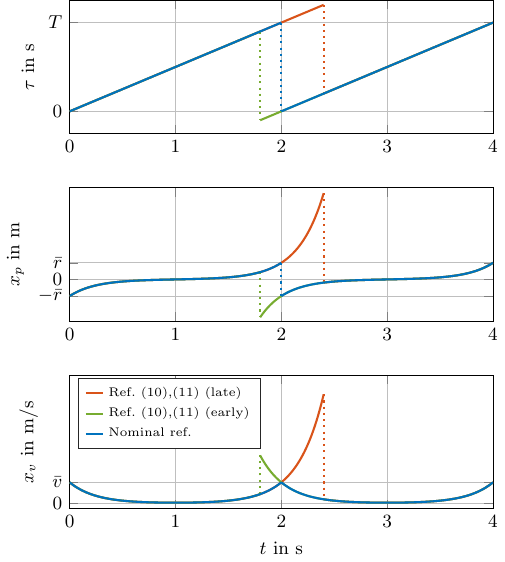}
    \caption{Comparison between the nominal periodic trajectory (blue) and the proposed parametrization $x_r(\tau)$ in \eqref{eq:reference_tau},\eqref{eq:tau_dynamics} when the robot is late (red) or early (green) with respect to the reference: timer state (top), position (middle) and velocity (bottom).}
    \label{fig:reference}
\end{figure}

To compute $T$ and simplify the reference expression, note that each step begins at $x_{r,0} := (-\bar r, \bar v)$.
Exploiting the fact that, for $A$ in \eqref{eq:com_ss}, we have
\begin{align}
    \label{eq:robot_exp}
    {\rm e}^{At} = 
    \begin{bmatrix}
        \cosh(\omega t) & \displaystyle \frac{ \sinh(\omega t) }{ \omega } \\[10pt]
        \omega \sinh(\omega t) & \cosh(\omega t)
    \end{bmatrix},
\end{align}
we may express the periodicity condition on $x_r$ as follows 
\begin{align}
\label{eq:periodicity}
    \bigmat{\bar r \\ \bar v} = 
    \exp{AT} x_{r,0} =     
    \exp{AT} \bigmat{-\bar r \\ \bar v},
\end{align}
which stems from the fact that the forward position changes sign across a foot swing, while the forward velocity remains the same.
It is straightforward to solve these equations, fixing the values for two of the three parameters $\bar r, \bar v, T$, to find a unique solution for the remaining quantity. 
In particular, the following (equivalent) relations hold for the periodic motion parameters
\begin{align}
    \label{eq:T}
    \bar v={\frac {\cosh \left( \omega\,T \right) + 1}{\sinh \left( \omega\,T \right)}}\omega \bar r_, \quad\quad
    T = \frac{1}{\omega} \ln \left( \displaystyle\frac{\bar v/\omega + \bar r}{\bar v/\omega - \bar r} \right).
\end{align}
Starting from the nominal periodic reference motion presented above, we propose the following parametrization for the reference trajectory
\begin{align}
\label{eq:reference_tau}
    x_r(\tau) = \exp{A \tau} x_{r,0},
\end{align}
where 
the hybrid dynamics of $\tau$ is characterized as follows
\begin{subequations}
    \label{eq:tau_dynamics}
    \begin{align}
        &\dot \tau = 1, &x_p = [1 \; 0] x \leq \bar r, \label{eq:tau_flow} \\
        &\tau^+ = \tau - T, &x_p = [1 \; 0] x = \bar r. \label{eq:tau_jump}
    \end{align}
\end{subequations}
Note that the jump dynamics of $\tau$ in \eqref{eq:tau_jump} is triggered by the value of the longitudinal position of the center of mass $x_p$, rather than the value of the timer itself.
This strategy is inspired by the well-known reference spreading technique (see, e.g., \cite{VanSteenACC23,BiemondTAC19,ForniTAC13} and references therein), but has the peculiar feature of characterizing a jump in the position rather than a jump in the velocity.
Due to the reference spreading mechanism, the proposed parametrization $x_r(\tau)$ in \eqref{eq:reference_tau} drifts away from the nominal periodic reference for a certain (finite) amount of time, if the robot is not synchronized with the reference.
However, it is immediate to verify that if $\tau \in [T, 2T]$ when $x \in D$, then our parametrization jumps back to the nominal reference, since $\tau^+ \in [0, T]$, see the example in Figure~\ref{fig:reference}.
Moreover, we show in the following that, when $x \in D$ in \eqref{eq:flow_jump_set}, the difference $T - \tau$ is an increasing function of the position error.
This will allow us to prove that, in view of an accurate choice of the input $u$, the timer $\tau$ asymptotically approaches the interval $[0,T]$ if the initial synchronization error is within a suitable estimate of the basin of attraction.
\subsection{Complete dynamics and error dynamics}
\label{sec:robot_error}
Augmenting the robot dynamics in \eqref{eq:com_ss} with the timer dynamics in \eqref{eq:tau_dynamics}, we obtain the following complete dynamics for the system 
\begin{subequations}
\label{eq:dyn_modified}
\begin{align}
    &\begin{bmatrix}
        \dot x \\ \dot \tau
    \end{bmatrix} = f(x,\tau) := \begin{bmatrix}
        A x + B u \\ 1
    \end{bmatrix}, &(x, \tau) \in \mathcal C \times [-T, 2T], \\
    &\begin{bmatrix}
        x^+ \\ \tau^+
    \end{bmatrix} = g(x, \tau) := \begin{bmatrix}
        x + F \\ \tau - T
    \end{bmatrix}, &(x, \tau) \in \mathcal D \times [-T, 2T],
\end{align}
\end{subequations}
where the reference trajectory is still parameterized as in \eqref{eq:reference_tau}.
In \eqref{eq:dyn_modified} we constrain the timer $\tau$ to belong to the compact set $[-T,2T]$ because of the local nature of the stabilization problem that we are studying.
Indeed, since we are dealing with an unstable continuous-time dynamics stabilized by a saturated control action (see, e.g., \cite{Sontag84}), which can only yield local results, a synchronization error of more than a step (i.e., $\tau < -T$ or $\tau > 2T$) is unlikely to be stabilizable.
Also note that the hybrid dynamics \eqref{eq:dyn_modified} satisfies the hybrid basic conditions \cite[Assumption 6.5]{TeelBook12}, since both $f$ and $g$ are continuous functions, and both $\mathcal C$ and $\mathcal D$ are closed subsets of $\mathcal X \times [-T, 2T]$.

Based on the hybrid dynamics \eqref{eq:dyn_modified} and on $x_r(\tau)$ in \eqref{eq:reference_tau}, we can define the tracking error as
\begin{align}
\label{eq:robot_error}
    \varepsilon = \left( \varepsilon_p, \, \varepsilon_v \right) := x - x_r(\tau).
\end{align}
To obtain the error dynamics, we need a preliminary result, that relates the position error $\varepsilon_p$ at jump times with the timer offset due to the reference spreading mechanism
\begin{align}
    \label{eq:tau_eps}
    \tau_\varepsilon := T - \tau.
\end{align}
In particular, we show next that, when $x \in D$,
\begin{align}
\label{eq:eta}
    \tau_\varepsilon &= \frac{1}{\omega} \ln \left( \eta(\varepsilon_p) \right) \\
    \eta \left( \varepsilon_p \right) &:= \displaystyle \frac{\varepsilon_p - \bar r + \sqrt{\varepsilon_p^2 - 2 \bar r \varepsilon_p + \left(\bar v /\omega \right)^2}}{\bar v / \omega - \bar r}, \nonumber
\end{align}
and that such an expression of $\tau_{\varepsilon}$ 
is an extended class $\mathcal K_\infty$ function of $\varepsilon_p$, in the sense that it is continuous, strictly increasing, zero at zero and positively and negatively unbounded, which generalizes the {\em extended class ${\mathcal K}$} concept introduced in \cite{AmesTabuadaTAC17}, to a globally invertible function.

\begin{proposition}
    \label{prop:kinf}
    Solutions to \eqref{eq:dyn_modified} satisfy \eqref{eq:eta} when $x \in D$.
    Moreover, $\bar v/\omega - \bar r > 0$ and the expression of $\tau_\varepsilon$ in \eqref{eq:eta} is an extended class $\mathcal K_\infty$ function of $\varepsilon_p$. Equivalently, $\eta(0)=1$, $\eta$ is strictly increasing, $\lim\nolimits\limits_{\varepsilon_p \rightarrow +\infty} \eta(\varepsilon_p) = +\infty$ and $\lim\nolimits\limits_{\varepsilon_p \rightarrow -\infty} \eta(\varepsilon_p) = 0$.
\end{proposition}

\begin{proof}
    From the definition of the error in \eqref{eq:robot_error}, we have
    \begin{align}
        \label{eq:position_error}
        \varepsilon_p = \bigmat{ 1 & 0 } \; (x - x_r).
    \end{align}
    In view of \eqref{eq:flow_jump_set}, \eqref{eq:robot_exp}, \eqref{eq:periodicity}, \eqref{eq:reference_tau}, and using the definition of $\tau_\varepsilon$ in \eqref{eq:tau_eps}, when $x \in D$, \eqref{eq:position_error} corresponds to
    \begin{align*}
        \varepsilon_p & = \bar r - \bigmat{ 1 & 0 } \;  \exp{A \tau} \bigmat{-\bar r \\ \bar v} \\
        &= \bar r - \bigmat{ 1 & 0 } \exp{- A \tau_\varepsilon} \bigmat{\bar r \\ \bar v}. \\
        &= \bar r - \cosh(-\omega \tau_\varepsilon) \bar r - \sinh(-\omega \tau_\varepsilon) \frac{\bar v}{\omega} \\
        &= \bar r - \frac{\exp{\omega \tau_\varepsilon}}{2} \left( \bar r - \frac{\bar v}{\omega} \right) - \frac{\exp{-\omega \tau_\varepsilon}}{2} \left( \bar r + \frac{\bar v}{\omega} \right).
    \end{align*}
    Rearranging the terms and multiplying by $\exp{\omega \tau_\varepsilon} > 0$ we obtain
    \begin{align}
        \label{eq:tau_eps_eqn}
        \left( \frac{\bar v}{\omega} - \bar r \right) \exp{2 \omega \tau_\varepsilon} - 2 (\varepsilon_p - \bar r) \exp{\omega \tau_\varepsilon} - \left( \bar r + \frac{\bar v}{\omega} \right) = 0.
    \end{align}
    From the expression of $\bar v$ in \eqref{eq:T} we obtain $\bar v / \omega - \bar r > 0$, so that the second order polynomial \eqref{eq:tau_eps_eqn} in $\exp{\omega \tau_\varepsilon}$ has two solutions of opposite sign (Descarte's rule of signs), where the negative one should be discarded because $\exp{\omega \tau_\varepsilon}>0$.
    Therefore, the unique positive solution to \eqref{eq:tau_eps_eqn} is $\exp{\omega \tau_\varepsilon} = \eta(\varepsilon_p)$, namely \eqref{eq:eta}.

    Let us now prove that $\varepsilon_p \mapsto \ln \left( \eta (\varepsilon_p) \right)$ is an extended class ${\mathcal K}_\infty$ function. Equivalently, we prove the four properties of $\eta$ stated in the proposition.
    
    It is immediate to verify by substitution that $\eta(0) = 1$.
    To verify that $\eta(\cdot)$ is strictly increasing, let us compute the derivative of its numerator with respect to $\varepsilon_p$ (recall that the denominator $\bar v/\omega - \bar r$ is a positive constant):
    \begin{align*}
        \left( \bar v / \omega - \bar r \right) \frac{\partial \eta}{\partial \varepsilon_p} = 1 + \frac{\varepsilon_p - \bar r}{\sqrt{\varepsilon_p^2 - 2 \bar r \varepsilon_p + (\bar v/\omega)^2}}.
    \end{align*}
    We show below that the norm of the last term in the expression above is smaller than 1. Indeed,
    \begin{align*}
        \left| \varepsilon_p - \bar r \right| &= \sqrt{\varepsilon_p^2 - 2 \bar r \varepsilon_p + \bar r^2 - (\bar v/\omega)^2 + (\bar v/\omega)^2} \\
        &< \sqrt{\varepsilon_p^2 - 2 \bar r \varepsilon_p + (\bar v/\omega)^2},
    \end{align*}
    where we used again the fact that $\bar v / \omega - \bar r > 0$.
    Thus, $\partial \eta/\partial \varepsilon_p > 0$ for any $\varepsilon_p$.
    Proving $\lim_{\varepsilon_p \rightarrow + \infty} \eta(\varepsilon_p) = + \infty$ is trivial by comparing the growth rates of the terms under the square root.
    Lastly, to compute $\lim_{\varepsilon_p \rightarrow - \infty} \eta(\varepsilon_p)$, let us define $\gamma := (\bar v/\omega)^2 - \bar r^2 > 0$, to simplify the notation.
    Then, to compute the limit of the numerator of $\eta(\cdot)$
    \begin{align*}
        &\lim\nolimits\limits_{\varepsilon_p \rightarrow - \infty} \varepsilon_p - \bar r + \sqrt{(\varepsilon_p-\bar r)^2 + \gamma},
    \end{align*}
    we divide and multiply by $\varepsilon_p - \bar r - \sqrt{(\varepsilon_p-\bar r)^2 + \gamma} < 0$, obtaining
    \begin{align*}
        \lim\nolimits\limits_{\varepsilon_p \rightarrow - \infty} &\frac{(\varepsilon_p-\bar r)^2 - \left( (\varepsilon_p-\bar r)^2 + \gamma \right)}{\varepsilon_p - \bar r - \sqrt{(\varepsilon_p-\bar r)^2 + \gamma}} \\
        &= \lim\nolimits\limits_{\varepsilon_p \rightarrow - \infty} \frac{-\gamma}{\varepsilon_p - \bar r - \sqrt{(\varepsilon_p-\bar r)^2 + \gamma}} = 0^+,
    \end{align*}
    which proves the same result for $\eta(\cdot)$ (since its denominator is positive), thus concluding the proof.
\end{proof}

Using Proposition~\ref{prop:kinf}, we obtain the following hybrid dynamics for the error coordinates
\begin{subequations}
    \label{eq:robot_error_dyn}
    \begin{align}
        \dot \varepsilon &= A \varepsilon + B u, \label{eq:robot_error_flow}  \\
        \varepsilon^+ &= \varepsilon + \begin{bmatrix} \delta_1\left( \varepsilon_p \right) \\ \delta_2 \left(\varepsilon_p \right) \end{bmatrix} := \varepsilon + \bar r \begin{bmatrix}
            \eta(\varepsilon_p) + \frac{1}{\eta(\varepsilon_p)} - 2 \\
            \omega \left( \frac{1}{\eta(\varepsilon_p)} - \eta(\varepsilon_p) \right)
        \end{bmatrix}, \label{eq:robot_error_jump}
    \end{align}
\end{subequations}
as stated in the next lemma, proved in Appendix~\ref{app:proof_error_dyn} to avoid breaking the flow of the exposition.
\begin{lemma}
    \label{lem:robot_error_dyn}
    In view of dynamics \eqref{eq:dyn_modified} and the reference parametrization \eqref{eq:reference_tau}, the tracking error $\varepsilon$ defined in \eqref{eq:robot_error} evolves according to the hybrid dynamics \eqref{eq:robot_error_dyn}.
\end{lemma}

The flow dynamics~\eqref{eq:robot_error_flow} is linear, and can be stabilized by an appropriate choice of a (linear) feedback control law, which would ensure a decrease of the error during the flow. However, the jump dynamics~\eqref{eq:robot_error_jump} is nonlinear, autonomous, and could in general result in an increase of the error. Given the strongly nonlinear nature of the jump dynamics, we are interested in obtaining a (conservative) bound on $|\delta_1|, |\delta_2|$ that is linear in $|\varepsilon_p|$.
This will allow us to use linear Lyapunov techniques and linear programming to analyze the stability of the closed-loop system in the next sections.
In particular, we show next that the following holds
\begin{align}
\label{eq:delta_bounds}
        \left| \delta_1 \left( \varepsilon_p \right) \right| \leq \frac{2 \xi}{\omega} |\varepsilon_p|, \quad\quad
        \left| \delta_2 \left( \varepsilon_p \right) \right| \leq 2 \xi |\varepsilon_p|,
\end{align}
where the scalar $\xi$, defined as
\begin{align}
\label{eq:xi}
    \xi &:= \frac{\bar r \omega}{\bar v/\omega-\bar r},
\end{align}
is positive because its denominator is positive by Proposition~\ref{prop:kinf}.
Additionally, we are interested in comparing the value of the tracking error before and after the jump, in view of a rescaling given by an exponential.
Namely, introducing $\delta_\alpha(\varepsilon_p) := \exp{-2 \alpha T} \varepsilon_p^+ - \varepsilon_p$ we can show that
\begin{align}
    \label{eq:delta_alpha}
    \begin{split}
        \left| \delta_\alpha \left( \varepsilon_p \right) \right| &:= \left| \exp{-2 \alpha T} \varepsilon_p^+ - \varepsilon_p \right| \\
        &\leq (1 - \exp{-(\omega + 2\alpha) T}) \left| \varepsilon_p \right|, \quad \forall \alpha > \omega.
    \end{split}
\end{align}
The rationale behind this bound will be clear in the proof of our main result.
Indeed, using a typical Lyapunov technique in the context of hybrid dynamical systems, we show that a quadratic Lyapunov function of the tracking error decreases with rate $\alpha$ along flows of the solution to the hybrid dynamics (namely, we will introduce a multiplicative diverging exponential $\exp{2\alpha \tau}$ in the Lyapunov function), so that the uncontrollable effect of the jumps can be compensated for.
Then, when comparing the value of the Lyapunov function before and after a jump, we may exploit the decrease of the timer state, as for \eqref{eq:delta_alpha}.

The three bounds stated above are summarized in the next lemma, proven in Appendix~\ref{app:proof_bounds} to avoid breaking the flow of the exposition.

\begin{lemma}
    \label{lem:bounds}
    The quantities $\delta_1 \left( \varepsilon_p \right), \delta_2 \left( \varepsilon_p \right)$ in \eqref{eq:robot_error_jump} and $\delta_\alpha \left( \varepsilon_p \right)$ satisfy \eqref{eq:delta_bounds}-\eqref{eq:delta_alpha}.
\end{lemma}

\section{Feedback Control Selection}
\label{sec:controller}

    We select the controller as a linear feedback with a symmetric saturation level, given by half of the foot length.
    In order to avoid windup phenomena, we also introduce a static linear antiwindup \cite{AW_book}, namely
    \begin{align}
        \label{eq:feedback}
        &u := \sat_{\bar u} (K \varepsilon + L/(1-L) \dz_{\bar u}(K \varepsilon)),
    \end{align}
    where $\varepsilon$ is defined in \eqref{eq:robot_error} and, for a generic $s \in \real$,
    \begin{align*}
        &\sat_{\bar u} (s) := \min \{ \max\{ s, - \bar u \}, \bar u \}, \\
        &\dz_{\bar u}(s) := s - \sat_{\bar u}(s).
    \end{align*}
    Inspired by \cite[Proposition 1]{BraunLCSS22}, for any given scalar $\alpha \in \real_{>0}$, the gains $L, K$ in \eqref{eq:feedback} are selected as
        \begin{align}
        \label{eq:gains}
            K = W Q^{-1}, \quad \quad L = X/U,
        \end{align}
        with matrix $Q=Q^\top := \smallmat{q_{11} & q_{12} \\ q_{12} & q_{22}} \in \real^{2 \times 2}$, row  vectors $W,Y \in \mathbb{R}^{2}$ and scalars $U, X \in \mathbb{R}$ solving the following convex optimization problem
        \begin{align}
        \label{eq:LMIs}
            \max\nolimits\limits_{Q,U,W,X,Y} &\log \mathrm{det} (Q) \;\;
            \mbox{subject to } \\ &\begin{array}{l}
                Q > 0, \; U > 0, \; \Delta(Q) < 0, \\
                \mathrm{He} \begin{bmatrix}
                    \alpha Q + AQ + BW & B(X-U) \\ W+Y & X-U
                \end{bmatrix} < 0, \\
                \begin{bmatrix}
                    \bar u^2 & Y \\ Y^\top & Q
                \end{bmatrix} \geq 0,
            \end{array} \nonumber
        \end{align}
        where $\Delta$ is a homogeneous function of $Q$, defined as $\Delta(Q) := \smallmat{\Delta_{11} & \Delta_{12} \\ \Delta_{12} & \Delta_{22}}$, with
        \begin{align}
        \label{eq:Delta}
            \Delta_{11} &:= \left( \exp{2(\omega-\alpha) T} - 1 \right) q_{22} + 4 \exp{-2 \alpha T} \xi \left(\xi q_{11} - \exp{\omega T} q_{12} \right), \nonumber \\
            \Delta_{12} &:= 2 \exp{-2 \alpha T} \xi q_{11} + \left( \exp{-(\omega + 2\alpha) T}- 1 \right) q_{12}, \\
            \Delta_{22} &:= \left( \exp{-2 \alpha T}-1 \right) q_{11}, \nonumber
        \end{align}
        and $\xi$ defined in \eqref{eq:xi}.
        We will prove the effectiveness of selection \eqref{eq:gains}-\eqref{eq:Delta} in the next section, when stating our main result, while for now the following proposition provides a useful guideline for selecting the parameter $\alpha$ and accurate feasibility conditions for \eqref{eq:gains}-\eqref{eq:Delta}.
    \begin{proposition}
        \label{claim:nec_suf}
        The LMI constraints \eqref{eq:gains}-\eqref{eq:Delta} are feasible if and only if $\alpha > \omega$.
    \end{proposition}
    \begin{proof}
        First we show that $\alpha > \omega$ is sufficient to find a feasible solution to \eqref{eq:gains}-\eqref{eq:Delta}.
        To this end, we select $X = 0$ and $Y = -W = -KQ$ for a matrix $K$ to be selected next.
        Then, the last two inequality constraints in \eqref{eq:LMIs} become
        \begin{align}
        \label{eq:useful1}
            \begin{split}
                &\begin{bmatrix}
                    \mathrm{He}((\alpha I + A + BK)Q) & -BU \\ -(BU)^\top & -2U
                \end{bmatrix} < 0, \\
                &\begin{bmatrix}
                    \bar u^2 & KQ \\ QK^\top & Q
                \end{bmatrix} \geq 0.
            \end{split}
        \end{align}
        The first inequality in \eqref{eq:useful1} is satisfied by selecting $U>0$ sufficiently small and
        \begin{align}
            \label{eq:R}
            R := \mathrm{He} ((\alpha I + A + BK)Q) < 0,
        \end{align}
        while the second inequality in \eqref{eq:useful1} is satisfied by selecting $Q$ with sufficiently small norm.
        Note that requiring $Q$ to have a small enough norm does not impose extra constraints.
        Indeed, all the inequalities left to be checked (i.e., $R < 0$ and $\Delta(Q) < 0$) are homogeneous functions of $Q$.

        We select now any $q_{11} > 0$ and $q_{12} = -2 \alpha q_{11}$.
        Then, we select $q_{22}$ large enough so that $\Delta(Q) < 0$ and $q_{22} > 4 \alpha^2 q_{11}$ both hold.
        Notice that, with fixed $q_{11}, q_{12}$ and $\alpha > \omega$, $\Delta(Q) < 0$ is verified by any $q_{22}$ sufficiently large, so the proposed selection is always possible.

        Lastly, we select the controller gain as $K = \left[ k_1 \;\; k_2 \right] = \omega^{-2} \left[ \bar k_1 + \omega^2 \;\; \bar k_2 \right]$, with
        \begin{align}
        \label{eq:ksel}
                \bar k_1 := \frac{q_{22}}{q_{11}} + \frac{4 \alpha^2 q_{22}}{q_{22} - 4 \alpha^2 q_{11}}, \;
                \bar k_2 := 2 \alpha + \frac{2 \alpha q_{22}}{q_{22} - 4 \alpha^2 q_{11}},
        \end{align}
        which are both positive because $q_{22} > 4 \alpha^2 q_{11}$.
        Let us now explicitly compute the elements of $R$ in \eqref{eq:R} with $A$ and $B$ as in \eqref{eq:com_ss}:
        \begin{align}
        \label{eq:useful2}
            R &= \mathrm{He} \left( \begin{bmatrix}
                \alpha & 1 \\
                \omega^2(1-k_1) & \alpha-\omega^2 k_2
            \end{bmatrix} \begin{bmatrix}
                q_{11} & q_{12} \\ q_{12} & q_{22}
            \end{bmatrix} \right) \\
            &= \begin{bmatrix}
                2(q_{12} + \alpha q_{11}) & q_{22} - \bar k_1 q_{11} + (2 \alpha - \bar k_2) q_{12} \\
                \star & -2 \bar k_1 q_{12} + 2 (\alpha - \bar k_2) q_{22}
            \end{bmatrix}. \nonumber
        \end{align}
        Substituting selections \eqref{eq:ksel} and $q_{12} = - 2 \alpha q_{11}$, \eqref{eq:useful2} yields
        \begin{align*}
            R \hspace{-1pt} &= \hspace{-4pt} \begin{bmatrix}
                \hspace{-2pt} -2 \alpha q_{11} & - \frac{4 \alpha^2 q_{11} q_{22}}{q_{22} - 4 \alpha^2 q_{11}} + 2 \alpha q_{11} \frac{2 \alpha q_{22}}{q_{22}- 4 \alpha^2 q_{11}} \\
                \star & 4 \alpha \hspace{-2pt} \left( \hspace{-2pt} q_{22} + \frac{4 \alpha^2 q_{11} q_{22}}{q_{22} - 4 \alpha^2 q_{11}} \hspace{-2pt} \right) \hspace{-4pt} - \hspace{-3pt} 2 \hspace{-2pt} \left( 
 \hspace{-2pt} \alpha \hspace{-2pt} + \hspace{-2pt} \frac{2 \alpha q_{22}}{q_{22} - 4 \alpha^2 q_{11}} \hspace{-2pt} \right) \hspace{-2pt} q_{22}
            \end{bmatrix} \\
            &= \hspace{-4pt} \begin{bmatrix}
                \hspace{-2pt} -2 \alpha q_{11} & 0 \\
                \star & \frac{2 \alpha q_{22}^2 - 8 \alpha^3 q_{11} q_{22} + 16 \alpha^3 q_{11} q_{22} - 4 \alpha q_{22}^2}{q_{22} - 4 \alpha^2 q_{11}}
            \end{bmatrix} \\
            &= \hspace{-4pt} \begin{bmatrix}
                \hspace{-2pt} -2 \alpha q_{11} & 0 \\
                0 & -2 \alpha q_{22}
            \end{bmatrix} < 0,
        \end{align*}
        therefore all the inequalities are satisfied.

        To show that $\alpha > \omega$ is necessary for \eqref{eq:gains}-\eqref{eq:Delta} to be feasible, we recall that a necessary condition for feasibility is negativity of the top-left term of $R$ in \eqref{eq:useful2}.
        Since $q_{11} > 0$ ($Q$ is positive definite), this implies that $q_{12} < 0$ for any feasible solution.
        Another necessary condition for the feasibility of \eqref{eq:gains}-\eqref{eq:Delta} is $\Delta_{11} < 0$ in \eqref{eq:Delta}: since $q_{22} > 0$ as well, the right-hand side of $\Delta_{11}$ has a negative term only if $\exp{2(\omega-\alpha)T} - 1 < 0$, i.e. $\alpha > \omega$.
        This proves that the condition is also necessary for the feasibility of \eqref{eq:gains}-\eqref{eq:Delta}, thus concluding the proof.
    \end{proof}

\section{Main stability result}
\label{sec:proof}
    Using the feedback controller \eqref{eq:gains}-\eqref{eq:Delta} in Section~\ref{sec:controller}, we show in this section that the compact set
    \begin{align}
        \label{eq:robot_A}
        \mathcal A \hspace{-1pt} := \hspace{-1pt} \left\{ (x, \tau) \in \mathcal X \times [-T,2T] \hspace{-1pt} : \hspace{-1pt} x = x_r(\tau), \tau \in [0, T] \right\},
    \end{align}
    where $x_r(\tau)$ is defined in \eqref{eq:reference_tau}, is locally asymptotically stable (LAS) for the error dynamics \eqref{eq:robot_error_dyn}, as stated in the next theorem.

    \begin{thm}
        \label{th:main}
        Consider dynamics \eqref{eq:dyn_modified}, where the control input is selected as in \eqref{eq:feedback}.
        If the control gains are chosen according to \eqref{eq:gains}-\eqref{eq:Delta} for some $\alpha > 0$, then the set $\mathcal A$ in \eqref{eq:robot_A} is LAS, and the set
        \begin{align}
            \label{eq:BoA_estimate}
            \begin{array}{l}
                \mathcal E := \hspace{-1pt} \big\{(x, \tau) \in \mathcal X \times [-T, 2T] : \vspace{5pt} \\
                \hspace{50pt} \left( x - x_r(\tau) \right)^\top Q^{-1} \hspace{-1pt} \left( x - x_r(\tau) \right) \leq 1 \big\}
            \end{array}
        \end{align}
        is contained in the basin of attraction of $\mathcal A$.
    \end{thm}
    \begin{proof}
        We consider the Lyapunov function candidate
        \begin{align}
            \label{eq:robot_V}
            V(x,\tau) := \exp{2 \alpha \tau} \left( x - x_r(\tau) \right)^\top P \left( x - x_r(\tau) \right) = \exp{2 \alpha \tau} \varepsilon^\top P \varepsilon,
        \end{align}
        where $P := Q^{-1} > 0$.
        
        To prove the theorem, we use a local version of \cite[Theorem 20]{GoebelCSM09}.
        We begin by noting that, as highlighted in Section~\ref{sec:robot_error}, the proposed hybrid dynamics satisfies the hybrid basic conditions.
        Moreover, $\mathcal A$ is compact, and $$\mathcal A \cap \mathcal D = \{ (x,\tau) : x_p = \bar r, x_v = \bar v, \tau = T \}.$$
        Then, the first assumption of \cite[Theorem 20]{GoebelCSM09} is proved by
        \begin{align}
            \label{eq:first}
            g(\mathcal A \cap \mathcal D) = \{ (x,\tau) : x_p = -\bar r, x_v = \bar v, \tau = 0 \} \in \mathcal A.
        \end{align}
        First we show that $V(x,\tau)=0$ if and only if $(x,\tau) \in \mathcal A$.
        To this end, note that $V(x,\tau) = 0$ if and only if $(x,\tau) \in \mathcal A_0$, with
        \begin{align}
            \label{eq:E0}
            \hspace{-3pt} \mathcal A_0 \hspace{-2pt} := \hspace{-2pt} \left\{ (x, \tau) \hspace{-2pt} \in \hspace{-1pt} \mathcal X \hspace{-1pt} \times \hspace{-1pt} [-T,2T] \hspace{-1pt} : \hspace{-1pt} \varepsilon = x - x_r(\tau) = 0 \right\},
        \end{align}
        where we could have $\tau \notin [0,T]$, due to the reference spreading mechanism.
        We show below that $\mathcal A_0$ in \eqref{eq:E0} coincides with $\mathcal A$ in \eqref{eq:robot_A}.
        To this end, note that $(x,\tau) \in \mathcal A_0$ implies $x_p = \psi(\tau)$, with
        \begin{align*}
            \psi(\tau) := \smallmat{1 & 0} x_r(\tau) = - \cosh \left( \omega \tau \right) \bar r + \frac{\sinh \left( \omega \tau \right)}{\omega} \bar v,
        \end{align*}
        where we used \eqref{eq:robot_exp} and \eqref{eq:reference_tau}.
        From \eqref{eq:robot_exp},\eqref{eq:periodicity} we also have that $\psi$ satisfies
        \begin{gather*}
            \psi(0) = - \bar r, \quad
            \psi(T) = \bar r, \\
            \frac{\partial \psi}{\partial \tau} = \exp{\omega T} (\omega \bar r + \bar v) + \exp{- \omega T} (\bar v - \omega \bar r) > 0,
        \end{gather*}
        where we used $\bar v - \omega \bar r > 0$, as established in Proposition~\ref{prop:kinf}.
        From the inverse function theorem, $\psi$ is an invertible function, and its inverse maps the iterval $[-\bar r, \bar r]$ to the interval $[0,T]$.
        Using this fact, together with the fact that $x \in \mathcal X \implies x_p \in [-\bar r, \bar r]$, we can say that $x_p = \psi(\tau)$ if and only if
        $$\tau = \psi^{-1} \left( x_p \right) \in [0,T],$$
        thus proving that $x = x_r(\tau)$ and $x \in \mathcal X$ implies $\tau \in [0,T]$ and therefore $\mathcal A_0 = \mathcal A$, namely $V$ in \eqref{eq:robot_V} is a valid Lyapunov function candidate for establishing LAS of $\mathcal A$.
        
        Consider now the directional derivative $\dot V$ of $V$ along flowing solutions: in view of \cite[Proposition 1]{BraunLCSS22}, if the first four inequalities in \eqref{eq:LMIs} hold, then $\varepsilon^\top P \varepsilon$ decreases exponentially with rate $2 \alpha$ along the flowing solutions to \eqref{eq:robot_error_dyn}, if the initial condition is inside the ellipsoidal estimate $\{ \varepsilon \in \real^2 : \varepsilon^\top P \varepsilon \leq 1\}$ in \eqref{eq:BoA_estimate} of the basin of attraction.
        That is, in our case, for \eqref{eq:robot_V}, $\dot V(x,\tau) < 0$ for all $(x,\tau)$ in $\mathcal E$ as defined in \eqref{eq:BoA_estimate}.

        Concerning the variation $\Delta V$ of $V$ across jumps from $D$, we have
        \begin{align*}
            \Delta V(x,\tau) &= \exp{2 \alpha \tau} \left(\exp{-2 \alpha T} \left(\varepsilon^+\right)^\top P \varepsilon^+ - \varepsilon^\top P \varepsilon \right) \\
            &= \exp{2 \alpha \tau}/\mathrm{det}(Q) \left(\exp{-2 \alpha T} \left(\varepsilon^+\right)^\top \tilde Q \varepsilon^+ - \varepsilon^\top \tilde Q \varepsilon \right),
        \end{align*}
        where
        $\tilde Q := \smallmat{q_{22} & \star \\ -q_{12} & q_{11}}$
        is the adjugate matrix of $Q = P^{-1}$, so that $\tilde Q / \mathrm{det} (Q) = P$.
        Since we are only interested in the sign of $\Delta V(x,\tau)$, we can neglect the positive scaling factor $\exp{2 \alpha \tau}/\mathrm{det}(Q)$ for brevity, and consider the following bounds, where we use \eqref{eq:robot_error_dyn},
        \begin{align}
        \label{eq:deltaV_1}
            \Delta \tilde V(x,\tau) &= \frac{\mathrm{det}(Q)}{\exp{2 \alpha \tau}} \Delta V(x,\tau) \nonumber \\
            &= q_{22} \left( \exp{-2 \alpha T} \left( \varepsilon_p^+ \right)^2 - \varepsilon_p^2 \right) \nonumber \\
                & \hspace{-40pt} - 2 q_{12} \left( \exp{-2 \alpha T} \varepsilon_p^+ \varepsilon_v^+ - \varepsilon_p \varepsilon_v \right) 
                + q_{11} \left( \exp{-2 \alpha T} \left( \varepsilon_v^+ \right)^2 - \varepsilon_v^2 \right) \nonumber \\
            &= q_{22} \left( \exp{-2 \alpha T} \left( \varepsilon_p^+ \right)^2 - \varepsilon_p^2 \right) \nonumber \\
                & \hspace{-40pt} - 2 q_{12} \left( \exp{-2 \alpha T} \varepsilon_p^+ \delta_2 +  \varepsilon_v \left( \exp{-2 \alpha T} \varepsilon_p^+ - \varepsilon_p \right) \right) \nonumber \\
                & \hspace{-40pt} + q_{11} \left( \left( \exp{-2 \alpha T} - 1 \right) \varepsilon_v^2 + 2 \exp{-2 \alpha T} \varepsilon_v \delta_2 + \exp{-2 \alpha T} \delta_2^2 \right) \nonumber \\
            &\leq q_{22} \left( \exp{-2 \alpha T} \left( \varepsilon_p^+ \right)^2 - \varepsilon_p^2 \right) \\
                & \hspace{-40pt} - 2 q_{12} \left( \exp{-2 \alpha T} \left| \varepsilon_p^+ \right| \left| \delta_2 \right| +  \left| \varepsilon_v \right| \left| \exp{-2 \alpha T} \varepsilon_p^+ - \varepsilon_p \right| \right) \nonumber \\
                & \hspace{-40pt} + q_{11} \left( \left( \exp{-2 \alpha T} - 1 \right) \varepsilon_v^2 + 2 \exp{-2 \alpha T} \left| \varepsilon_v \right| \left| \delta_2 \right| + \exp{-2 \alpha T} \delta_2^2 \right), \nonumber
        \end{align}
        where the last inequality holds in view of the fact that $q_{11}, q_{22} > 0$ and $q_{12}<0$, as discussed in the proof of Proposition~\ref{claim:nec_suf}.

        Now, using Lemma~\ref{lem:bounds}, we can derive the following bound for jumps induced by \eqref{eq:robot_error_jump}, where we also use \eqref{eq:T},
        \begin{align}
            \label{eq:robot_Another_bound}
            \begin{array}{rl}
                \left| \varepsilon_p^+ \right| \leq
                \left| \varepsilon_p \right| + \left| \delta_1 \left( \varepsilon_p \right) \right|
                &\leq \left( 1 + \frac{2 \bar r}{\bar v / \omega - \bar r} \right) \left| \varepsilon_p \right| \\
                &= \frac{\bar v / \omega + \bar r}{\bar v / \omega - \bar r} \left| \varepsilon_p \right|
                = \exp{\omega T} \left| \varepsilon_p \right|.
            \end{array}
        \end{align}
        Using \eqref{eq:delta_bounds}, \eqref{eq:delta_alpha} and \eqref{eq:robot_Another_bound} in \eqref{eq:deltaV_1}, we obtain, by using \eqref{eq:LMIs} in the last inequality,
        \begin{align*}
            \Delta \tilde V(x,\tau) &\leq q_{22} \left( \exp{2 (\omega - \alpha) T} - 1 \right) \varepsilon_p^2 \\
            & \hspace{-40pt} - 2 q_{12} \left( 2 \exp{(\omega - 2 \alpha) T} \xi \varepsilon_p^2 + \left( 1 - \exp{-(\omega + 2 \alpha) T} \right) \left| \varepsilon_p \right| \left| \varepsilon_v \right| \right) \\
            & \hspace{-40pt} + q_{11} \left( \left( \exp{-2 \alpha T} - 1 \right) \varepsilon_v^2 + 4 \exp{-2 \alpha T} \xi \left| \varepsilon_p \right| \left| \varepsilon_v \right| + 4 \exp{-2 \alpha T} \xi^2 \varepsilon_p^2 \right) \\
            &= \smallmat{\left| \varepsilon_p \right| & \left| \varepsilon_v \right|} \Delta(Q)
                \smallmat{\left| \varepsilon_p \right| \\ \left| \varepsilon_v \right|} < 0, \quad \forall (x,\tau) \in D \setminus \mathcal A,
        \end{align*}
        which implies negative definiteness of $\Delta V$.
        Note that we used \eqref{eq:delta_alpha}, which holds due to Lemma~\ref{lem:bounds}.

        By virtue of \eqref{eq:first} and due to the negative definiteness of $\dot V$ and $\Delta V$ established above for $(x,\tau) \in \mathcal E$, we can use a local version of \cite[Theorem 20]{GoebelCSM09} to conclude that the compact set $\mathcal A$ (where $V(x, \tau) = 0$) is LAS for the error dynamics, with basin of attraction including $\mathcal E$.
    \end{proof}

    Notice that, in view of the hybrid dynamics satisfying the hybrid basic conditions, and the consequent well-posedness established in \cite[Chapter 6]{TeelBook12}, the stability property proven in Theorem~\ref{th:main} is also robust, as discussed in \cite[Chapter 7]{TeelBook12}.
    This implies that bounded uncertainties in the center of mass position at the moment of the foot change do not compromise the stability of the tracking error.
    
\section{Simulations}
\label{sec:simulations}

    In this section we report the results of numerical simulations performed with the proposed control law \eqref{eq:feedback} with the LIPM model \eqref{eq:dyn_modified} (equivalently, \eqref{eq:robot_error_dyn}).
    The robot dynamics is obtained by selecting $z_c = \SI{0.58}{\meter}$, which yields $\omega = \sqrt{g/z_c} \approx 4.1~\mathrm{rad/s}$.
    The reference trajectory is obtained selecting $\bar r = \SI{0.15}{\meter}$ and $T = \SI{1.2}{\second}$, which yield $\bar v = \SI{0.552}{\meter/\second}$ from \eqref{eq:T}.
    The half-foot size, which defines the saturation limit, is selected as $\bar u = 7.5 \cdot 10^{-2} \SI{}{\meter}$.

    Using these parameters, and selecting $\alpha = 4.2 > \omega = 4.1$, the control gains computed by solving the optimization problem \eqref{eq:gains}-\eqref{eq:Delta} (whose feasibility is ensured by Proposition~\ref{claim:nec_suf}) are
    $$K = \bigmat{198.3 & 42.2}, \quad\quad L = 0.94$$
    Interestingly, these gains are relatively close to the optimal gains suggested in~\cite{Villa2017}, which for our system, considering a control period of 1 ms, would be $K = \bigmat{244.4 & 59.6}$.
    Our gains give the following matrix $P$, defining the estimated basin of attraction
    $$ P = 10^4 \cdot \bigmat{8.54 & 0.1 \\ 0.1 & 0.42}.$$ 
    \begin{figure}[hbt]
        \centering
        \includegraphics{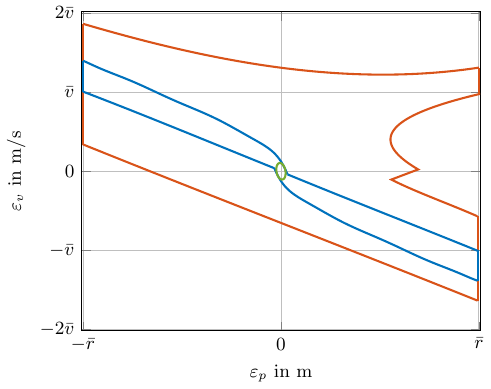}
        \caption{Comparison among the set $\mathcal E$ in \eqref{eq:BoA_estimate} (green), the set of initial conditions such that $\dot V$ is non increasing (blue) and the set of initial conditions such that the robot state converges to the periodic reference (red).}
        \label{fig:basin_of_attraction}
    \end{figure}
    \begin{figure}[hbt]
        \centering
        \includegraphics{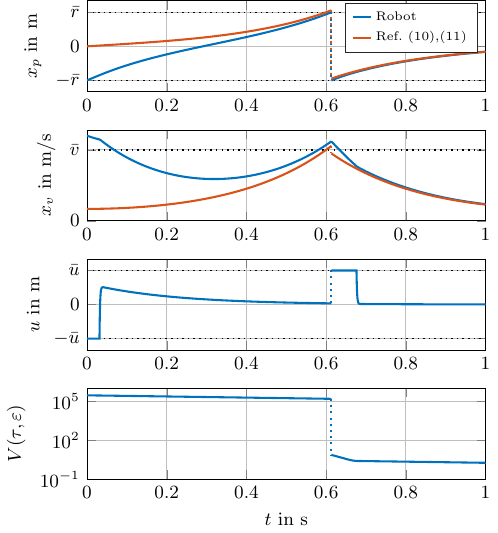}
        \caption{Time evolution of the robot's center of mass position (first plot) and velocity (second plot), compared to the reference trajectory \eqref{eq:reference_tau},\eqref{eq:tau_dynamics}. Corresponding input signal (third plot), where the black dotted lines characterize the saturation limits.
        Evolution of the Lyapunov function $V$ in \eqref{eq:robot_V} along the simulated response (fourth plot).}
        \label{fig:robot_trajectories}
    \end{figure}
    Figure~\ref{fig:basin_of_attraction} reports a representation of the ellipsoidal estimate (depicted with a green contour).
    The ellipsoid is plotted in error coordinates, so that it can be easily visualized in a 2D plot.
    The figure also shows the set of initial values of the error coordinates such that the Lyapunov function is strictly decreasing along flows and across jumps of the solution (blue contour).
    This set was obtained through numerical integration of the dynamics, with the reference trajectory initialized at $\tau(0,0) = T/2$, to be able to represent the set with a 2D plot.
    Lastly, Figure~\ref{fig:basin_of_attraction} depicts a numerical estimate of the basin of attraction of \eqref{eq:robot_A} (red contour), again obtained from numerical integration with $\tau(0,0) = T/2$.
    While the estimate $\mathcal E$ in \eqref{eq:BoA_estimate} obtained from the LMI formulation \eqref{eq:LMIs} is not very extended if compared with the other sets, we can see that the ellipsoid is almost tangent to the set corresponding to the decrease of $V$.
    This suggests that our estimate could be the best result achievable with convex analysis tools applied to the Lyapunov function $V$ in \eqref{eq:robot_V}.
    The numerical estimate of the basin of attraction, on the other hand, suggests that the selection of a different Lyapunov function, specifically tailored to this system, could produce less conservative results.
    This investigation is the objective of ongoing work.

    Figure~\ref{fig:robot_trajectories} shows the time evolution of the robot state, compared to the computed reference \eqref{eq:reference_tau},\eqref{eq:tau_dynamics}.
    Since the ellipsoidal estimate of the basin of attraction is not extended, we selected initial conditions in the blue region in Figure~\ref{fig:basin_of_attraction}.
    In particular, the solution was simulated selecting $x(0,0) = (-\bar r, 1.2 \bar v)$ and $\tau(0,0) = T/2$.
    From the plots, we can see that both the robot position and velocity correctly converge to the reference values.
    The figure also includes the corresponding evolution of the control input $u$ computed according to \eqref{eq:feedback} and the desirable decrease of the Lyapunov function $V$ in \eqref{eq:robot_V} along the simulated response.

\section{Full humanoid robot simulations}
\label{sec:full_body_sim}
\begin{figure}[tb]
    \centering
    \includegraphics[width=0.55\columnwidth]{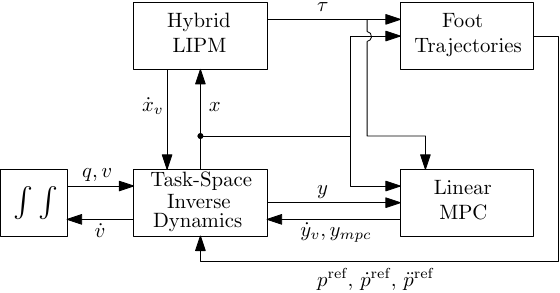}
    \caption{General control framework for the full-body simulations.}
    \label{fig:framework}
\end{figure}

We evaluate here the proposed control law in a suitable whole-body bipedal walking model. The overall simulation architecture is depicted in Fig.~\ref{fig:framework}. Since the hybrid LIPM control law~\eqref{eq:feedback} only stabilizes the longitudinal CoM motion, we introduce a standard Model Predictive Controller (block ``Linear MPC'', Section~\ref{ssec:linear_mpc}) for the lateral direction. Moreover, since the CoM and the CoP described by the LIP model are not sufficient to characterize the whole motion of a biped robot, we introduce the ``Foot Trajectories'' block (Section~\ref{ssec:foot_traj}), which computes the reference motion for both feet. Finally, the desired CoM, CoP and foot trajectories are mapped to joint torques using the state-of-the-art Task-Space Inverse Dynamics approach (TSID, Section~\ref{ssec:tsid}). 

The simulation architecture has been implemented in Python, using the C++ library TSID~\cite{DelPrete2015tsid} and the multi-body dynamics C++ library Pinocchio~\cite{Carpentier2019}. 
The control framework runs with a sampling time $t_s = 1$ ms, except for the linear MPC block, which uses a higher sampling time $t_{\text{MPC}} = 30$ ms, because of computational requirements. Animation videos of the simulation results can be seen at \href{https://youtu.be/l42ksZsonTo}{\texttt{https://youtu.be/l42ksZsonTo}}.

    \subsection{Linear Model Predictive Control}
    \label{ssec:linear_mpc}
    To stabilize the lateral CoM state $y \coloneqq (y_p,\, y_v)$ we use a standard Model Predictive Control (MPC) formulation~\cite{Diedam2008,HerdtAR10} based on the LIPM dynamics~\eqref{eq:com_ss}. 
    This formulation avoids the hybrid nature of the system by relying on pre-defined footstep positions and durations. 
    In particular, we assume that the robot walks in the forward direction, with no lateral displacement except for the unavoidable zero-average lateral CoM oscillations.
    Therefore, defining the half lateral distance between the feet as $\bar y$, we set the reference footstep position equal to $\bar{y}$ for the left foot, and $-\bar y$ for the right foot.

    Regarding the footstep durations, in classical approaches~\cite{Diedam2008,HerdtAR10} they are set to a constant value $T$.
    However, in our model, the footstep duration depends on the longitudinal CoM state $x$, thus it can differ from the nominal duration $T$ in \eqref{eq:T}. 
    To account for this, at each MPC iteration we compute the duration of the current footstep by simulating the error dynamics~\eqref{eq:robot_error_dyn} with the controller~\eqref{eq:feedback}, and evaluating the residual time $t_{\text{res}}$ needed to finish the step. 
    The duration of the remaining footsteps included in the MPC horizon, which we choose to be $2 T$, is simply set to the nominal value $T$ because we observed that adapting it would have negligible effects.
    
    The MPC problem is formulated as follows, where the generic sampled value $\zeta[k]$ denotes $\zeta(k \, t_{\text{MPC}})$:
    \begin{equation}
        \begin{aligned}
        \min_{ Y_j, U_j} \, & \sum_{k = j}^{j+N-1}  \gamma \norm{y_{v}[k] }^2 + \alpha \norm{u_{y}[k] - z_{y}[k]}^2    \\
        \text{s.t.} \quad &  y[k+1] =  A_d \,y[k] + B_d \, u_{y}[k],  \\ 
        & z_{y}[k] - \frac{w_y}{2} \le u_{y}[k] \le z_{y}[k] + \frac{w_y}{2}, \\
        & y[j] = \hat{y}[j], \qquad \forall \  k \in \{j, \dots, j+N-1\}.
    \end{aligned}
    \label{eq:lmpc}
    \end{equation}
    where $Y_j = \{y[j],\dots, y[j+N]\}$ and $U_j = \{u_{y}[j],\dots, u_{y}[j+N-1]\}$ are the lateral CoM states and CoP positions in the considered $N$-sample horizon, and $\hat{y}[j]$ is the current lateral CoM state. Moreover, $z_{y}$ denotes the lateral reference CoP, which corresponds to the center of the pre-defined footstep position, $w_y$ is the foot width, $(A_d,\, B_d)$ are the discretized versions of the state and input matrices $(A, B)$ in~\eqref{eq:flow_law} with sample time $t_{\text{MPC}} > 0$, and $\alpha, \gamma$ are user-defined positive values that weigh the relative importance of the three cost terms. 
    Problem~\eqref{eq:lmpc} can be solved as a Quadratic Program with off-the-shelf solvers.

    \subsection{Swing Foot Trajectories}
    \label{ssec:foot_traj}
    Another crucial part of bipedal walking is the motion of the feet, which is often represented using polynomials or splines~\cite{Leziart2021}. 
    Defining the 3D foot position as $p(t) \coloneqq (p_x(t),\, p_y(t),\, p_z(t))$, we know that the swing foot trajectory must satisfy the following nominal conditions:
    \begin{equation}
    \begin{aligned}
    p_{x}(0) &= p_{x,\text{init}}, \qquad &p_{x}(T) &= p_{x,\text{final}}, \\
    p_{y}(0) &= p_{y,\text{init}}, \qquad &p_{y}(T) &= p_{y,\text{final}}, \\
    p_{z}(0) &= p_{z}(T) = 0, \qquad &p_z(T/2) &= p_z^{\text{max}},\\
    \dot p(0) &= \dot p(T) = 0, \qquad   &\ddot p(0) &= \ddot p(T) = 0,
    \end{aligned}
    \label{eq:poly_conditions}
    \end{equation}
    with $T$ defined in \eqref{eq:T}, $p_{x,\text{init}}, p_{y, \text{init}}$ and $p_{x,\text{final}}, p_{y,\text{final}}$ denote the initial and final $x$-$y$ swing foot position, while $p_z^{\text{max}}$ denotes the maximum foot height reached during the swing.
    We used 5-th order polynomials for the $x$ and $y$ directions and 6-th order polynomials for the $z$ direction. This is because, as shown in~\eqref{eq:poly_conditions}, the vertical foot trajectory has the additional constraint of reaching a predefined height at $t=T/2$.

    As stated above, the footstep duration is not known a priori, thus the reference swing foot trajectories must be adapted using the following time-scaling factor:
    \begin{equation}
        \phi = \frac{T}{t+t_{\text{res}}},
    \end{equation}
    where $t$ is the time elapsed since the beginning of the current footstep, and $t_{\text{res}}$ is the residual time required to finish the step, computed from \eqref{eq:robot_error_dyn},\eqref{eq:feedback} with the approach described in Section~\ref{ssec:linear_mpc}. 
    The reference swing foot position $p^{\text{ref}}$, velocity $\dot p^{\text{ref}}$ and acceleration $\ddot p^{\text{ref}}$, to be delivered to the TSID block in Figure~\ref{fig:framework}, are then computed by evaluating the polynomials at $\phi \, t$, and applying the appropriate scaling:
    \begin{equation}
        p^{\text{ref}}(t) = p(\phi\, t), \quad \dot p^{\text{ref}}(t) = \phi\, \dot p(\phi\, t), \quad \ddot p^{\text{ref}}(t) = \phi^2 \ddot p(\phi \, t).
    \end{equation}
    
    \subsection{Task-Space Inverse Dynamics}
    \label{ssec:tsid}
    To map the reference CoM and foot trajectories to joint torques, we need to use the nonlinear robot dynamics. 
    Denoting with $q \in \mathbb{SE}(3) \times \real^{n}$ and $v \in \real^{6+n}$, respectively, the position and the velocity of the base (which is the floating body of the robot, defined in an Euclidean group $\mathbb{SE}(3)$) and the $n$ joints, the equations of motion of a legged robot are \cite[Chapter 3]{SicilianoKhatib2008}:
    \begin{equation}
        M(q) \dot v + b(q,v) = S^\top \tau_j + J_c(q)^\top \lambda,
        \label{eq:robot_dyn}
    \end{equation}
    where $M(q) \in \real^{(6+n)\times(6+n)}$ is the mass matrix, $b(q,v) \in \real^{6+n}$ are the generalized torques due to the nonlinear effects (Coriolis, centrifugal and gravity effects), $\tau_j \in \real^n$ are the joint torques, $S \in \real^{n\times (6+n)}$ is the joint-selection matrix, $J_c(q) \in \real^{k \times (6+n)}$ is the contact Jacobian and $\lambda \in \real^k$ is the contact force vector.  
    If the robot is in rigid contact with the environment, as a biped robot with one foot on the ground, the dynamics must also consider the contact constraints.
    Since the contact points $c(q) \in \real^{k}$ do not move (unless the contact is broken), their velocities $\dot{c}$ and accelerations $\ddot{c}$ are zero.
    The contact constraints are usually accounted for at the acceleration level, so that they are linear with respect to $\dot v$:
    \begin{equation}
        \ddot{c} = J_c(q) \dot{v} + \dot{J}_c(q, v) v = 0
        \label{eq:contact_constraints}
    \end{equation}
    Starting from the reference CoM and the foot trajectories, we use the reactive control framework Task-Space Inverse Dynamics (TSID)~\cite{DelPrete2015tsid} to generate joint torques that track the reference trajectories. 
    This framework acts similarly to an input-output feedback linearization, but relies on the formulation of a QP to compute the joint torques. 
    The QP uses the joint accelerations $\dot v$, the joint torques $\tau_j$ and the contact forces $\lambda$ as decision variables.
    The system dynamics~\eqref{eq:robot_dyn} appear in the problem as equality constraints, while tracking performance metrics are used as cost function. 
    In particular, given an output function $o_i(q)$ that should track a reference trajectory $o_i^r(t)$ (e.g., for the CoM or the swing foot), we can differentiate $o_i$ twice to find its relationship to $\dot{v}$:
    \begin{equation}
        \ddot o_{i} = J_i(q) \dot v + \dot J_i(q, v) v,
        \label{eq:acc_constraint}
    \end{equation}
    where $J_i(q) = \partial o_i / \partial q$ is the task Jacobian. 
    We can then compute a desired value for $\ddot o_i$ by using a stabilizing proportional-derivative feedback term:
    \begin{equation}
        \ddot o_i^d = \ddot o_i^r + k_i^d (\dot o_i^r - \dot o_i) + k_i^p (o_i^r - o_i), 
        \label{eq:ee_pd}
    \end{equation}
    with $k_i^p>0, k_i^d>0$ being the proportional and derivative gains of the $i$-th task. 
    Then, the tracking performance can be measured by computing the difference between the desired and the real value of $\ddot o_i$ in \eqref{eq:ee_pd} and \eqref{eq:acc_constraint}.
    Several tasks can be accounted for at the same time by simply adding these tracking performance metrics, weighted by user-defined penalties $w_i > 0$. 
    Then the task-space control law can be computed by solving the following QP:
    \begin{equation}
        \begin{aligned}
            \min_{\tau_j ,\dot v, \lambda} \quad & \sum_{i} w_i \norm{J_i(q) \dot v + \dot J_i (q, v) v - \ddot o_i^d}^2 \\
            \text{s.t.} \quad & M(q) \dot v + b(q, v) = S^\top \tau_j + J_c(q)^\top \lambda \\
            & J_c(q) \dot{v} + \dot{J}_c(q,v) v = 0
        \end{aligned}
        \label{eq:tsid}
    \end{equation}
    In addition to the CoM and the swinging foot motion tasks, we considered the following tasks: {\it i)} horizontal torso orientation, {\it ii)} joint posture and {\it iii)} force regularization to maintain the contact forces close to the center of the friction cones.

    \subsection{Simulation overview}

    Given the starting footstep sequence, we derive the initial state of the feet using the approach in Section~\ref{ssec:foot_traj}. 
    The initial joint configuration is computed by solving the Inverse Geometry optimization problem with a numerical solver
    to match the desired positions of the CoM and the feet. 
    Similarly, the initial velocities are computed by solving a linear system of equations to match the initial velocities of the CoM and the feet.



    The CoM state and the foot trajectories computed by the ``Hybrid LIMP'', ``Linear MPC'' and ``Foot Trajectories'' blocks are used as references for problem \eqref{eq:tsid}. 
    Since the LIPM does not represent the vertical CoM direction, we simply consider a constant CoM height $z_p^{\text{ref}}$.
    Finally, the robot state is computed by double integration of the accelerations $\dot v$ computed by TSID, via the Explicit Euler scheme.  
    
    \subsection{Simulation Results}
    \begin{table*}[t]
        \centering
        \caption{Main parameters used in the full-body simulations, divided into their respective subcategories.}
        \resizebox{\textwidth}{!}{
        \begin{tabular}{lccccccccccccccc}
            \toprule
            & \multicolumn{3}{c}{\textbf{Hybrid LIPM}} & \multicolumn{5}{c}{\textbf{MPC}} & \multicolumn{7}{c}{\textbf{TSID}}  \\
            \cmidrule(lr){2-4} \cmidrule(lr){5-9} \cmidrule(lr){10-16}
            \textbf{Param} & $t_s$ (\SI{}{\milli\second}) & $T$ (\SI{}{\second}) & $\bar r$ (\SI{}{\meter}) & $\alpha$ &  $\gamma$ & $t_{\text{MPC}}$ (\SI{}{\milli\second}) & $w_y$ (\SI{}{\meter})& $\bar y$ (\SI{}{\meter}) & &\emph{CoM} & \emph{Torso} & \emph{Foot} & \emph{Contact} & \emph{Pos} & \emph{Reg} \\
            \cmidrule(lr){11-16}
            \multirow{2}{*}{\textbf{Value}} & \multirow{2}{*}{1} & \multirow{2}{*}{1.2} & \multirow{2}{*}{0.1} & \multirow{2}{*}{10}&   \multirow{2}{*}{0.01} & \multirow{2}{*}{30}&  \multirow{2}{*}{0.1} & \multirow{2}{*}{0.096} & \textbf{Weight} & 1 & 0.01 & 1& $10^4$ & $10^{-4}$ & $10^{-5}$ \\
            &  & & & & & & & & \textbf{Gain} & 10 & 10 & 10 & 5 & 1\\
            \bottomrule
        \end{tabular}
        }
        \label{tab:simu_params}
    \end{table*}
    This subsection reports the simulation results on the 37 degree-of-freedom humanoid robot Romeo. Table~\ref{tab:simu_params} shows the simulation parameters. The rows \emph{Weight} and \emph{Gain} refer to the weights $w_i$ and proportional gains $k^p_i$ associated with the TSID tasks. To obtain critical damping, we have set each derivative gain as $k_i^d = 2 \sqrt{k_i^p}$.
    \begin{figure}[hbt]
        \centering
        \includegraphics[width=0.55\columnwidth]{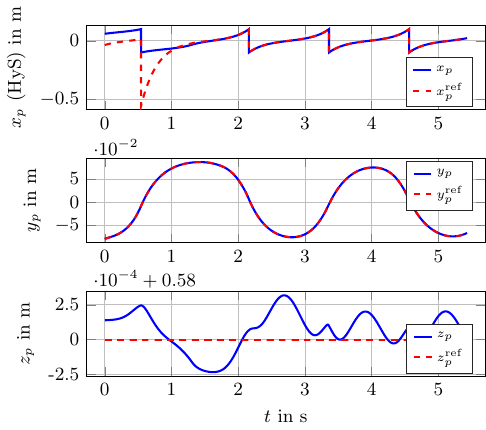}
        \caption{Real and reference CoM position.}
        \label{fig:com_pos}
    \end{figure}

    \begin{figure}[hbt]
        \centering
        \includegraphics[width=0.55\columnwidth]{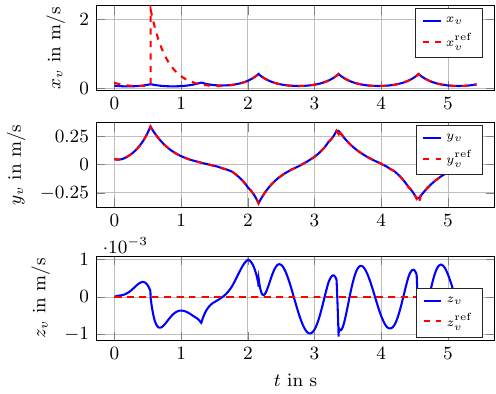}
        \caption{CoM velocity in the $x$ (top), $y$ (middle) and $z$ (bottom) directions.}
        \label{fig:com_vel}
    \end{figure}

    Figure~\ref{fig:com_pos} shows the CoM position trajectories during four bipedal footsteps. The initial state of the CoM is $x(0) = (\SI{0.06}{\meter},\, \SI{0.08}{\meter/\second})$ and $y(0) = (\SI{-0.08}{\meter},\, \SI{0.05}{\meter/\second})$. The top plot reports the $x$ coordinate in the Hybrid System reference frame. 
    Initially, the robot is early with respect to the reference trajectory \eqref{eq:reference_tau},\eqref{eq:tau_dynamics} (red dashed curve). As a consequence, the position error $\varepsilon_p$ defined in \eqref{eq:robot_error} sensibly increases across the first  jump. The feedback law \eqref{eq:feedback} then correctly decreases the robot velocity, as shown in Figure~\ref{fig:com_vel}, reporting the time evolution of the CoM velocities. This corrective feedback action results in a footstep duration higher than the nominal value $T$, and we can see that this is correctly handled in the lateral direction by the MPC stabilization described in Section~\ref{ssec:linear_mpc}.
    \begin{figure}[tb]
        \centering
        \includegraphics[width=0.55\columnwidth]{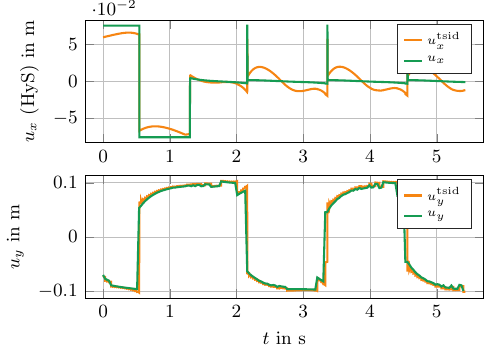}
        \caption{Comparison of the CoP positions obtained by the Hybrid System/Lateral MPC versus TSID.}
        \label{fig:cop}
    \end{figure}

    To analyze the similarity between the LIPM and the full biped model, we plot in Figure~\ref{fig:cop} the evolution of the CoP. The green lines represent the CoP computed with the hybrid model ($x$ axis) and the MPC ($y$ axis), while the orange lines represent the CoP of the full-body robot, that we computed from the contact forces $\lambda$ optimized by TSID. The two quantities are close, in particular for the lateral direction. In the $x$ direction, there are some deviations due to the nonlinear effects of the full model (e.g. angular momentum variations, which are supposed null in the LIPM), but the response follows the ideal CoP from the hybrid LIPM.
    \begin{figure}[tb]
        \centering
        \includegraphics[width=0.55\columnwidth]{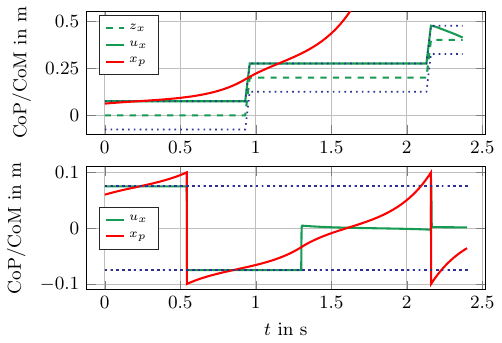}
        \caption{Comparison between the first step obtained using classic MPC (top) and our hybrid model (bottom), along the $x$ direction. The green dashed line is the center of the stance foot, while the blue dotted lines are the foot borders.}
        \label{fig:mpc_vs_hs}
    \end{figure}

    Figure~\ref{fig:mpc_vs_hs} shows a comparison between our method and the classic MPC approach \cite{Diedam2008,HerdtAR10} (without footstep optimization) for both the $x$-$y$ directions. In the beginning (until $\approx$0.55 s) both approaches behave in the same way, setting the CoP to the tip of the foot to decelerate the CoM as much as possible. However, when the CoM reaches 0.1 m, the hybrid model triggers an early footstep. The new footstep position allows the robot to set the CoP in front of the CoM, therefore decelerating it. In robotics this behavior is called ``capturing''~\cite{Koolen2012} the diverging CoM. The classic MPC approach instead waits until the end of the nominal footstep duration $T$ to finish the footstep (at $\approx$0.9 s). This leads to a high CoM velocity, which can no longer be recovered, leading to a diverging CoM.

\section{Conclusions}
    We proposed a novel formulation for the problem of trajectory tracking for a walking robot.
    The introduction of a hybrid dynamics allowed moving the reference frame to the center of the foot on the ground at every step of the robot, thus ensuring that the center of mass position coordinate evolves in a compact set.
    By parametrizing the reference trajectory with a timer state, we obtained the hybrid error dynamics, and using LMI design tools we provided a tuning procedure for the feedback controller gains ensuring local asymptotic stability of the synchronization set.


    The longitudinal hybrid model, together with a state-of-the-art MPC controller for the lateral CoM motion, has been used as reference for whole-body biped simulations through the TSID framework. The results have shown the applicability of our approach to periodic locomotion, and the robustness with respect to the classic MPC with open loop CoP reference. 

    Future work includes extensions of our theory to the lateral motion, as well as the investigation of more general Lyapunov functions and nonlinear stabilizers, in addition to performing experiments on a real humanoid robot.

    \bibliographystyle{unsrt}
    \bibliography{refs.bib}

\begin{thebibliography}{10}

\bibitem{Wieber2016}
Pierre-Brice Wieber, Russ Tedrake, and Scott Kuindersma.
\newblock {\em Modeling and Control of Legged Robots}, pages 1203--1234.
\newblock Springer International Publishing, Cham, 2016.

\bibitem{RoboticSurvey}
Patrick~M. Wensing, Michael Posa, Yue Hu, Adrien Escande, Nicolas Mansard, and
  Andrea Del~Prete.
\newblock Optimization-based control for dynamic legged robots, 2022.

\bibitem{KajitaICRA03}
S.~Kajita, F.~Kanehiro, K.~Kaneko, K.~Fujiwara, K.~Harada, K.~Yokoi, and
  H.~Hirukawa.
\newblock Biped walking pattern generation by using preview control of
  zero-moment point.
\newblock In {\em IEEE International Conference on Robotics and Automation},
  volume~2, pages 1620--1626, 2003.

\bibitem{HerdtAR10}
Andrei Herdt, Holger Diedam, Pierre-Brice Wieber, Dimitar Dimitrov, Katja
  Mombaur, and Moritz Diehl.
\newblock Online walking motion generation with automatic footstep placement.
\newblock {\em Advanced Robotics}, 24(5-6):719--737, 2010.

\bibitem{PontonICRA18}
Brahayam Ponton, Alexander Herzog, Andrea Del~Prete, Stefan Schaal, and Ludovic
  Righetti.
\newblock On time optimization of centroidal momentum dynamics.
\newblock In {\em 2018 IEEE International Conference on Robotics and Automation
  (ICRA)}, pages 5776--5782, 2018.

\bibitem{GallowayAccess15}
Kevin Galloway, Koushil Sreenath, Aaron~D. Ames, and Jessy~W. Grizzle.
\newblock Torque saturation in bipedal robotic walking through control lyapunov
  function-based quadratic programs.
\newblock {\em IEEE Access}, 3:323--332, 2015.

\bibitem{AmesACC18}
Shishir Kolathaya, Jacob Reher, Ayonga Hereid, and Aaron~D. Ames.
\newblock Input to state stabilizing control lyapunov functions for robust
  bipedal robotic locomotion.
\newblock In {\em 2018 Annual American Control Conference (ACC)}, pages
  2224--2230, 2018.

\bibitem{AmesRAL21}
Wen-Loong Ma, Noel Csomay-Shanklin, Shishir Kolathaya, Kaveh~Akbari Hamed, and
  Aaron~D. Ames.
\newblock Coupled control lyapunov functions for interconnected systems, with
  application to quadrupedal locomotion.
\newblock {\em IEEE Robotics and Automation Letters}, 6(2):3761--3768, 2021.

\bibitem{RinconTSMCS22}
Karla Rincon, Isaac Chairez, and Wen Yu.
\newblock Finite-time output feedback robust controller based on tangent
  barrier lyapunov function for restricted state space for biped robot.
\newblock {\em IEEE Transactions on Systems, Man, and Cybernetics: Systems},
  52(8):5042--5055, 2022.

\bibitem{AmesACC20}
Jenna Reher, Claudia Kann, and Aaron~D. Ames.
\newblock An inverse dynamics approach to control lyapunov functions.
\newblock In {\em 2020 American Control Conference (ACC)}, pages 2444--2451,
  2020.

\bibitem{AmesTAC14}
Aaron~D. Ames, Kevin Galloway, Koushil Sreenath, and Jessy~W. Grizzle.
\newblock Rapidly exponentially stabilizing control lyapunov functions and
  hybrid zero dynamics.
\newblock {\em IEEE Transactions on Automatic Control}, 59(4):876--891, 2014.

\bibitem{AmesNAHS17}
Shishir Kolathaya and Aaron~D. Ames.
\newblock Parameter to state stability of control lyapunov functions for hybrid
  system models of robots.
\newblock {\em Nonlinear Analysis: Hybrid Systems}, 25:174--191, 2017.

\bibitem{SidorovNAHS19}
Eric Sidorov and Miriam Zacksenhouse.
\newblock Lyapunov based estimation of the basin of attraction of poincare maps
  with applications to limit cycle walking.
\newblock {\em Nonlinear Analysis: Hybrid Systems}, 33:179--194, 2019.

\bibitem{MorrisTAC09}
Benjamin Morris and Jessy~W. Grizzle.
\newblock Hybrid invariant manifolds in systems with impulse effects with
  application to periodic locomotion in bipedal robots.
\newblock {\em IEEE Transactions on Automatic Control}, 54(8):1751--1764, 2009.

\bibitem{GrizzleAuto14}
Jessy~W. Grizzle, Christine Chevallereau, Ryan~W. Sinnet, and Aaron~D. Ames.
\newblock Models, feedback control, and open problems of 3d bipedal robotic
  walking.
\newblock {\em Automatica}, 50(8):1955--1988, 2014.

\bibitem{TeelBook12}
Rafal Goebel, Ricardo Sanfelice, and Andrew Teel.
\newblock {\em Hybrid dynamical systems: Modeling, stability, and robustness}.
\newblock Springer, 2012.

\bibitem{VanSteenACC23}
Jari~J. van Steen, Nathan van~de Wouw, and Alessandro Saccon.
\newblock Robot control for simultaneous impact tasks through time-invariant
  reference spreading.
\newblock In {\em 2023 American Control Conference (ACC)}, pages 46--53, 2023.

\bibitem{BiemondTAC19}
Mark Rijnen, J.~J.~Benjamin Biemond, Nathan van~de Wouw, Alessandro Saccon, and
  Henk Nijmeijer.
\newblock Hybrid systems with state-triggered jumps: Sensitivity-based
  stability analysis with application to trajectory tracking.
\newblock {\em IEEE Transactions on Automatic Control}, 65(11):4568--4583,
  2020.

\bibitem{ForniTAC13}
Fulvio Forni, Andrew~R. Teel, and Luca Zaccarian.
\newblock Follow the bouncing ball: Global results on tracking and state
  estimation with impacts.
\newblock {\em IEEE Transactions on Automatic Control}, 58(6):1470--1485, 2013.

\bibitem{SacconICRA17}
Mark Rijnen, Eric de~Mooij, Silvio Traversaro, Francesco Nori, Nathan van~de
  Wouw, Alessandro Saccon, and Henk Nijmeijer.
\newblock Control of humanoid robot motions with impacts: Numerical experiments
  with reference spreading control.
\newblock In {\em 2017 IEEE International Conference on Robotics and Automation
  (ICRA)}, pages 4102--4107, 2017.

\bibitem{Sontag84}
{Eduardo D.} Sontag.
\newblock An algebraic approach to bounded controllability of linear
  systems†.
\newblock {\em International Journal of Control}, 39(1):181--188, 1984.

\bibitem{AmesTabuadaTAC17}
A.D. Ames, X.~Xu, J.W. Grizzle, and P.~Tabuada.
\newblock Control barrier function based quadratic programs for safety critical
  systems.
\newblock {\em IEEE Transactions on Automatic Control}, 62(8):3861--3876, 2017.

\bibitem{AW_book}
Luca Zaccarian and Andrew~R Teel.
\newblock {\em Modern anti-windup synthesis: control augmentation for actuator
  saturation}, volume~38.
\newblock Princeton University Press, 2011.

\bibitem{BraunLCSS22}
P.~Braun, G.~Giordano, C.~M. Kellett, and L.~Zaccarian.
\newblock An asymmetric stabilizer based on scheduling shifted coordinates for
  single-input linear systems with asymmetric saturation.
\newblock {\em IEEE Control Systems Letters}, 6:746--751, 2022.

\bibitem{GoebelCSM09}
Rafal Goebel, Ricardo~G. Sanfelice, and Andrew~R. Teel.
\newblock Hybrid dynamical systems.
\newblock {\em IEEE Control Systems Magazine}, 29(2):28--93, 2009.

\bibitem{Villa2017}
Nahuel~A. Villa and Pierre-Brice Wieber.
\newblock {Model Predictive Control of Biped Walking with Bounded
  Uncertainties}.
\newblock In {\em IEEE RAS International Conference on Humanoid Robots}, 2017.

\bibitem{DelPrete2015tsid}
Andrea~Del Prete, Francesco Nori, Giorgio Metta, and Lorenzo Natale.
\newblock Prioritized motion-force control of constrained fully-actuated
  robots: "task space inverse dynamics".
\newblock {\em Robotics and Autonomous Systems}, 63:150--157, 1 2015.

\bibitem{Carpentier2019}
Justin Carpentier, Guilhem Saurel, Gabriele Buondonno, Joseph Mirabel, Florent
  Lamiraux, Olivier Stasse, and Nicolas Mansard.
\newblock The pinocchio c++ library : A fast and flexible implementation of
  rigid body dynamics algorithms and their analytical derivatives.
\newblock In {\em 2019 IEEE/SICE International Symposium on System Integration
  (SII)}, pages 614--619, 2019.

\bibitem{Diedam2008}
Holger Diedam, Dimitar Dimitrov, Pierre~Brice Wieber, Katja Mombaur, and Moritz
  Diehl.
\newblock Online walking gait generation with adaptive foot positioning through
  linear model predictive control.
\newblock In {\em 2008 IEEE/RSJ International Conference on Intelligent Robots
  and Systems, IROS}, pages 1121--1126, 2008.

\bibitem{Leziart2021}
Pierre-Alexandre Léziart, Thomas Flayols, Felix Grimminger, Nicolas Mansard,
  and Philippe Souères.
\newblock Implementation of a reactive walking controller for the new
  open-hardware quadruped solo-12.
\newblock In {\em 2021 IEEE International Conference on Robotics and Automation
  (ICRA)}, pages 5007--5013, 2021.

\bibitem{SicilianoKhatib2008}
Bruno Siciliano and Oussama Khatib, editors.
\newblock {\em Springer Handbook of Robotics}.
\newblock Springer, Berlin, Heidelberg, 2008.

\bibitem{Koolen2012}
Twan Koolen, T.~de~Boer, J.~Rebula, Ambarish Goswami, and Jerry Pratt.
\newblock {Capturability-based analysis and control of legged locomotion, Part
  1: Theory and application to three simple gait models}.
\newblock {\em The International Journal of Robotics Research},
  31(9):1094--1113, jul 2012.

\end{thebibliography}

\appendix

\section{Proof of Lemma~\ref{lem:robot_error_dyn}}
\label{app:proof_error_dyn}
    We begin by noticing that the flow dynamics of the parametrization $x_r(\tau)$ is given by
    \begin{align*}
        \dot x_r = \frac{\partial x_r}{\partial \tau} \cdot \dot \tau = A \exp{A \tau} x_{r,0} \cdot 1 = A x_r.
    \end{align*}
    Thus, we get
    \begin{align*}
        \dot \varepsilon = \dot x - \dot x_r = Ax + Bu - A x_r = A \varepsilon + Bu.
    \end{align*}
    For the jump dynamics of the reference, we have
    \begin{align*}
        x_r^+ = x_r(\tau^+) = \exp{A(\tau-T)}x_{r,0} = \exp{-AT} x_r,
    \end{align*}
    therefore
    \begin{align}
        \label{eq:jump_proof1}
        \varepsilon^+ &= x^+ - x_r^+ = x - \bigmat{2 \bar r \\ 0} - \exp{-AT} x_r \nonumber \\
        &= \varepsilon + \bigmat{- \bar r \\ \bar v} - \bigmat{\bar r \\ \bar v} + \left( I - \exp{-AT} \right) x_r \nonumber \\
        &= \varepsilon + \left( \exp{-AT} - I \right) \bigmat{\bar r \\ \bar v} + \left( I - \exp{-AT} \right) \exp{-A \tau_\varepsilon} \bigmat{\bar r \\ \bar v} \nonumber \\
        &= \varepsilon - \left( I - \exp{-AT} \right) \left( I - \exp{-A \tau_\varepsilon} \right) \bigmat{\bar r \\ \bar v}.
    \end{align}
    Now, to simplify the following calculations, notice that
    \begin{align*}
        \exp{-At} = \frac{\exp{-\omega t}}{2} S_- + \frac{\exp{\omega t}}{2} S_+ := \frac{\exp{-\omega t}}{2} \smallmat{1 & \frac{1}{\omega} \\ \omega & 1} + \frac{\exp{\omega t}}{2} \smallmat{1 & -\frac{1}{\omega} \\ -\omega & 1},
    \end{align*}
    and it holds that
    \begin{align*}
        S_- S_+ = S_+ S_- = 0, \quad S_- S_- = 2 S_-, \quad S_+ S_+ = 2 S_+.
    \end{align*}
    Therefore, \eqref{eq:jump_proof1} becomes
    \begin{align}
        \label{eq:jump_proof2}
        \varepsilon^+ &= \varepsilon - \left( I - \frac{S_-}{2 \exp{\omega T}} - \frac{\exp{\omega T} S_+}{2} \right) \left( I - \frac{S_-}{2 \eta} - \frac{\eta S_+}{2} \right) \bigmat{\bar r \\ \bar v} \nonumber \\
        &= \varepsilon - \bigg( I + \left( \frac{\exp{-\omega T}- 1}{\eta} - \exp{-\omega T} \right) \frac{S_-}{2} \\
        & \hspace{90pt} + \left( \eta(\exp{\omega T}-1) - \exp{\omega T} \right) \frac{S_+}{2} \bigg) \bigmat{\bar r \\ \bar v}, \nonumber
    \end{align}
    where we removed the dependence of $\eta$ from $\varepsilon_p$, to avoid overloading the notation.
    Now notice that
    \begin{align}
        \label{eq:ff1}
        \begin{split}
            &S_- \bigmat{\bar r \\ \bar v} = \bigmat{1 \\ \omega}(\bar v/\omega + \bar r), \\
            &S_+ \bigmat{\bar r \\ \bar v} = \bigmat{-1 \\ \omega}(\bar v/\omega - \bar r).
        \end{split}
    \end{align}
    From this, recalling the expression for the reference period in \eqref{eq:T}, we get
    \begin{align}
        \label{eq:ff2}
        \begin{split}
            \exp{-\omega T} S_- \bigmat{\bar r \\ \bar v} &= \bigmat{1 \\ \omega}(\bar v/\omega - \bar r), \\
            \exp{\omega T} S_+ \bigmat{\bar r \\ \bar v} &= \bigmat{-1 \\ \omega}(\bar v/\omega + \bar r).
        \end{split}
    \end{align}
    Using \eqref{eq:ff1} and \eqref{eq:ff2}, \eqref{eq:jump_proof2} simplifies to
    \begin{align*}
        \varepsilon^+ &= \varepsilon - \bigg( \bigmat{ \bar r \\ \bar v } + \bigmat{1 \\ \omega} \left( - \frac{\bar r}{\eta} + \frac{\bar r - \bar v/\omega}{2} \right) \\
        & \hspace{100pt} + \bigmat{-1 \\ \omega} \left( \bar r \eta - \frac{\bar r + \bar v/\omega}{2} \right) \bigg) \\
        & = \varepsilon - \bigmat{\bar r - \frac{\bar r}{\eta} - \bar r \eta + \bar r \\ \bar v + \omega \left( \bar r \eta - \frac{\bar r}{\eta} - \bar v/\omega \right)},
    \end{align*}
    which is equivalent to \eqref{eq:robot_error_jump}, thus concluding the proof.

\section{Proof of Lemma~\ref{lem:bounds}}
\label{app:proof_bounds}
    First, we rationalize the expression of $1/\eta \left( \varepsilon_p \right)$, which gives
    \begin{align}
    \label{eq:eta_inv}
        \frac{1}{\eta \left( \varepsilon_p \right)} &= \displaystyle \frac{\bar v / \omega - \bar r}{\varepsilon_p - \bar r + \sqrt{\varepsilon_p^2 - 2 \bar r \varepsilon_p + \left(\bar v /\omega \right)^2}} \nonumber \\
        &= \displaystyle \frac{(\bar v / \omega - \bar r) \left(\varepsilon_p - \bar r - \sqrt{\varepsilon_p^2 - 2 \bar r \varepsilon_p + \left(\bar v /\omega \right)^2}\right)}{\bar r^2 - \left( \bar v / \omega \right)^2} \nonumber \\
        &= \frac{- \varepsilon_p + \bar r + \sqrt{\varepsilon_p^2 - 2 \bar r \varepsilon_p + \left(\bar v /\omega \right)^2}}{\bar v / \omega + \bar r}.
    \end{align}
    To simplify the computations, let us denote $\kappa := \sqrt{\varepsilon_p^2 - 2 \bar r \varepsilon_p + \left(\bar v /\omega \right)^2}$.
    Then, $\delta_1 \left( \varepsilon_p \right)$ and $\delta_2 \left( \varepsilon_p \right)$ in \eqref{eq:robot_error_jump} can be written as
    \begin{align*}
        \delta_1 \left( \varepsilon_p \right) &= \bar r \left( \frac{\varepsilon_p - \bar r + \kappa}{\bar v / \omega - \bar r} + \frac{- \varepsilon_p + \bar r + \kappa}{\bar v / \omega + \bar r} - 2 \right) \\
        &= \frac{2 \bar r}{\left( \bar v / \omega \right)^2 - \bar r^2} \left( \bar r \varepsilon_p + \bar v / \omega \kappa - \left( \bar v / \omega \right)^2 \right), \\[5pt]
        \delta_2 \left( \varepsilon_p \right) &= \bar r \omega \left( \frac{- \varepsilon_p + \bar r + \kappa}{\bar v / \omega + \bar r} - \frac{\varepsilon_p - \bar r + \kappa}{\bar v / \omega - \bar r} \right) \\
        &= \frac{2 \bar r \omega}{\left( \bar v / \omega \right)^2 - \bar r^2} \left( \bar v / \omega \left( \bar r - \varepsilon_p \right) - \bar r \kappa \right),
    \end{align*}
    and their derivatives with respect to $\varepsilon_p$ are
    \begin{align*}
        \begin{split}
            \frac{\partial \delta_1}{\partial \varepsilon_p} &= \frac{2 \bar r}{\left( \bar v / \omega \right)^2 - \bar r^2} \left( \bar r + \bar v / \omega \frac{\varepsilon_p - \bar r}{\kappa} \right), \\[5pt]
            \frac{\partial \delta_2}{\partial \varepsilon_p} &= \frac{2 \bar r \omega}{\left( \bar v / \omega \right)^2 - \bar r^2} \left( - \bar v / \omega + \bar r \frac{\bar r - \varepsilon_p}{\kappa} \right).
        \end{split}
    \end{align*}
    Using the fact that $|\left( \varepsilon_p - \bar r \right) / \kappa| \leq 1$ (shown in the proof of Proposition~\ref{prop:kinf}), after some algebraic derivations we can conclude that
    \begin{align}
        \label{eq:delta_derivatives}
        \begin{split}
            &- \frac{2 \bar r}{\bar v / \omega + \bar r} \leq \frac{\partial \delta_1}{\partial \varepsilon_p} \leq \frac{2 \bar r}{\bar v / \omega - \bar r} = \frac{2 \xi}{\omega}, \\
            & -\frac{2 \bar r \omega}{\bar v / \omega - \bar r} = - 2 \xi \leq \frac{\partial \delta_2}{\partial \varepsilon_p} \leq - \frac{2 \bar r \omega}{\bar v / \omega + \bar r},
        \end{split}
    \end{align}
    which, together with the fact that $\delta_1(0) = \delta_2(0) = 0$, proves \eqref{eq:delta_bounds}.

    Concerning expression \eqref{eq:delta_alpha}, we have
    \begin{align*}
        \delta_\alpha \left( \varepsilon_p \right) = 
        \left( \exp{-2 \alpha T} - 1 \right) \varepsilon_p + \exp{-2 \alpha T} \delta_1.
    \end{align*}
    The derivative of this quantity with respect to $\varepsilon_p$ is given by
    \begin{align*}
        \frac{\partial \delta_\alpha}{\partial \varepsilon_p} &= \left( \exp{-2 \alpha T} - 1 \right) + \exp{-2 \alpha T} \frac{\partial \delta_1}{\partial \varepsilon_p} \\
        &= \exp{-2 \alpha T} \left( 1 + \frac{\partial \delta_1}{\partial \varepsilon_p} \right) - 1.
    \end{align*}
    This expression is always negative if $\alpha > \omega$. Indeed, using \eqref{eq:delta_derivatives} we can write
    \begin{align*}
        \max \left\{ \frac{\partial \delta_\alpha}{\partial \varepsilon_p} \right\} &= \exp{-2 \alpha T} \left( 1 + \frac{2 \bar r}{\bar v / \omega - \bar r} \right) - 1 \\
        &= \exp{-2 \alpha T} \frac{\bar v / \omega + \bar r}{\bar v / \omega - \bar r} - 1 \\
        &= \exp{(-2 \alpha + \omega) T} - 1 < 0,
    \end{align*}
    where we applied the definition in \eqref{eq:T} in the last step.
    Then, the maximum of $\left| \partial \delta_\alpha/\partial \varepsilon_p \right|$ corresponds to
    \begin{align*}
        \max \left\{ \left| \frac{\partial \delta_\alpha}{\partial \varepsilon_p} \right| \right\} &= 1 - \exp{-2 \alpha T} \left( 1 - \frac{2 \bar r}{\bar v / \omega + \bar r} \right) \\
        &= 1 - \exp{-2 \alpha T} \frac{\bar v / \omega - \bar r}{\bar v / \omega + \bar r} = 1 - \exp{-(\omega + 2 \alpha) T},
    \end{align*}
    which, again, together with the fact that $\delta_\alpha(0) = 0$, proves \eqref{eq:delta_alpha}.

\end{document}